\documentclass[12pt]{article}
\pdfoutput=1
\usepackage[utf8]{inputenc}
\usepackage[DIV13]{typearea}
\usepackage{ragged2e}
\usepackage{calligra,amsmath,amsfonts,bbm,mathrsfs,amssymb}
\usepackage{slashed,cancel,units}
\usepackage[usenames,dvipsnames]{xcolor}
\usepackage{cite}
\usepackage{hyperref}
\hypersetup{colorlinks=true,urlcolor=Magenta,anchorcolor=blue,citecolor=blue,filecolor=blue,
            linkcolor=Magenta,menucolor=blue, linktocpage=true,pdfproducer=medialab}
\usepackage[english]{babel}
\usepackage{catchfile}
\usepackage{indentfirst}
\usepackage{graphicx}
\usepackage{float}
\usepackage{enumerate}
\usepackage{subcaption}
\usepackage{multirow}
\usepackage{makecell}
\usepackage{subcaption}
\usepackage{bm}

\newcommand{\orcid}[1]{\href{https://orcid.org/#1}{\includegraphics[width=8pt]{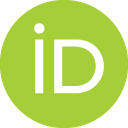}\hspace{2mm}\color{Blue}{orcid.org/#1}}}

\newcommand{\email}[1]{\href{mailto:#1}{\tt #1}}

\numberwithin{equation}{section}

\newcommand{\getenv}[2][]{%
  \CatchFileEdef{\temp}{"|kpsewhich --var-value #2"}{}%
  \if\relax\detokenize{#1}\relax\temp\else\let#1\temp\fi}
\getenv[\USER]{USER}
\newcommand{\red}[1]{\color{red} #1 \color{black}}
\newcommand{\blue}[1]{\color{blue} #1 \color{black}}

\newcommand{\be}{\begin{equation}}
\newcommand{\ee}{\end{equation}}
\newcommand{\ba} {\begin{equation}\begin{aligned}}
\newcommand{\ea} {\end{aligned}\end{equation}}
\newcommand{\bea}{\begin{eqnarray}}
\newcommand{\eea}{\end{eqnarray}}

\newcommand{\hc}{\text{h.c.}}
\newcommand{\ov}[1]{\overline{#1}}
\newcommand{\nn}{\nonumber}

\newcommand{\diag}{\mathtt{diag}}
\newcommand{\Tr}{\mathrm{Tr}}

\renewcommand{\Re}{\mathrm{Re}}
\renewcommand{\Im}{\mathrm{Im}}

\newcommand{\abs}[1]{\left| #1 \right|}
\def\Ds{\slashed{D}}
\newcommand{\absval}[1]{\left| #1 \right|} 

\def\MLsM{ML$\sigma$M }

\newcommand{\sL}{\mathscr{L}}
\newcommand{\sB}{ B}

\newcommand{\sT}{ T}

\newcommand{\nb}{\mathtt{b}}
\newcommand{\nB}{\mathtt{B}}
\newcommand{\nh}{\mathtt{h}}
\newcommand{\nK}{\mathtt{K}}
\newcommand{\nq}{\mathtt{q}}
\newcommand{\nQ}{\mathtt{Q}}
\newcommand{\ns}{\mathtt{s}}
\newcommand{\nt}{\mathtt{t}}
\newcommand{\nT}{\mathtt{T}}

\newcommand{\cB}{\mathcal{B}}

\newcommand{\cM}{\mathcal{M}}
\newcommand\cO{\mathcal{O}}

\newcommand{\cT}{\mathcal{T}}

\newcommand{\cZ}{\mathcal{Z}}

\newcommand{\wH}{\widetilde{H}}

\def\TeV{\text{ TeV}}
\def\GeV{\text{ GeV}}

\newcommand{\s}{\sigma}

\newcommand{\mo}{M_1}

\newcommand{\mf}{M_5}

\newcommand{\lamo}{\Lambda_1}

\newcommand{\lamt}{\Lambda_2}

\newcommand{\lamth}{\Lambda_3}

\newcommand{\mpf}{M'_5}

\newcommand{\lampo}{\Lambda'_1}

\newcommand{\lampt}{\Lambda'_2}

\newcommand{\lampth}{\Lambda'_3}

\newcommand{\yo}{\text{y}_1}
\newcommand{\yt}{\text{y}_2}

\newcommand{\mpo}{M'_1}

\newcommand{\ypo}{\text{y}'_1}
\newcommand{\ypt}{\text{y}'_2}

\begin{document}
\renewcommand*{\thefootnote}{\fnsymbol{footnote}}
\begin{titlepage}
\vspace*{-1cm}
\red{\flushleft{FTUAM-20-27}
\hfill{IFT-UAM/CSIC-20-167}
\hfill{ZU-TH-50/20}
}\\
\vskip 2cm
\begin{center}
\mathversion{bold}
\blue{{\LARGE\bf Probing Effective Field Theory Approach in}}\\[4mm] 
\blue{{\LARGE\bf the CP Violating Minimal Linear $\sigma$ Model}}\\[4mm]
\mathversion{normal}
\vskip .3cm
\end{center}
\vskip 0.5  cm

\begin{center}
{\large\bf J.~Alonso-Gonz\'alez}${}^{a)}$\footnote{\email{j.alonso.gonzalez@csic.es}; \orcid{0000-0002-0345-3860}},
{\large\bf  J.M.~Lizana}${}^{b)}$\footnote{\email{jlizana@physik.uzh.ch}; \orcid{0000-0002-2998-7158}}
{\large\bf  V.~Mart\'inez-Fern\'andez}${}^{c)}$\footnote{\email{Victor.Martinez-fernandez@ncbj.gov.pl}; \orcid{0000-0002-0581-7154}},\\[2mm]
{\large\bf  L.~Merlo}${}^{a)}$\footnote{\email{luca.merlo@uam.es}; \orcid{0000-0002-5876-4105}}, and
{\large\bf  S.~Pokorski}${}^{d)}$\footnote{\email{Stefan.Pokorski@fuw.edu.pl}; \orcid{0000-0002-3750-1330}}
\vskip .7cm
{\footnotesize
${}^{a)}$ Departamento de F\'isica Te\'orica and Instituto de F\'isica Te\'orica UAM/CSIC,\\
Universidad Aut\'onoma de Madrid, Cantoblanco, 28049, Madrid, Spain
\vskip .2cm
${}^{b)}$ Physik-Institut, Universität Zürich, CH-8057 Zürich, Switzerland
\vskip .2cm
${}^{c)}$ National Centre for Nuclear Research (NCBJ),\\ Pasteura 7, 02-093 Warsaw, Poland
\vskip .2cm
${}^{d)}$ Institute of Theoretical Physics, Faculty of Physics,\\ 
University of Warsaw, Pasteura 5, PL 02-093, Warsaw, Poland
}
\end{center}

\vskip 2cm
\begin{abstract}
\justify

The Minimal Linear $\sigma$ Model is a useful theoretical laboratory. One can investigate in a perturbative renormalisable model the properties of the Higgs boson as a pseudo-Goldstone boson, the phenomenological effects of the radial mode of the field $\ns$ which spontaneously breaks the global $SO(5)$ symmetry and the validity of conclusions based on the Effective Field Theory approach with the field $\ns$ in the spectrum, after the decoupling of heavy degrees of freedom. In this paper all those issues are discussed in the framework of the Minimal Linear $\s$ Model with CP violating phases leading to pseudoscalar components in the effective Standard Model Yukawa couplings. Also the character of the electroweak phase transition in the presence of the field $\ns$ is investigated.

\end{abstract}
\end{titlepage}
\setcounter{footnote}{0}

\pdfbookmark[1]{Table of Contents}{tableofcontents}
\tableofcontents

\renewcommand*{\thefootnote}{\arabic{footnote}}
\section{Introduction}
There has been some interest in perturbative models with the Higgs boson as a pseudo-Goldstone boson from an extended symmetry, both non-supersymmetric~\cite{Georgi:1974yw,Georgi:1975tz,ArkaniHamed:2001nc,Schmaltz:2005ky} and supersymmetric~\cite{Birkedal:2004xi,Chankowski:2004mq,Berezhiani:2005pb,Falkowski:2006qq,Chang:2006ra,Craig:2013fga,Katz:2016wtw}. Apart from investigating the Higgs boson properties in renormalisable beyond the Standard Model (BSM) scenarios, their simplest versions permit for studying the phenomenological role of an additional electroweak (EW) singlet $\ns$ whose vacuum expectation value (VEV) $v_\ns$ breaks spontaneously the extended global symmetry. The physical scalar spectrum consists of two mass eigenstates (we denote them by $h$ and $\s$) of the mass matrix which mixes the radial mode of $\ns$ and the neutral component of the Higgs doublet (it is natural to assume $v_\ns>v_\nh$ where $v_\nh$ is the VEV of the Higgs doublet). Since the presence of the scalar $\sigma$ in the spectrum restores perturbative unitarity and assuming that the other degrees of freedom are heavier, one can discuss such perturbative scenarios in the effective field theory (EFT) approach. Besides those from the SMEFT~\cite{Buchmuller:1985jz,Grzadkowski:2010es}, it has to include operators, beginning with dim 5, that are built out of the scalar $\ns$ and the SM fields. In such a model independent approach it is interesting to investigate the potential phenomenological role of the additional to SMEFT operators whose contribution depends on $v_\ns>v_\nh$.\footnote{There is a vast literature in various contexts on the models with additional scalar electroweak singlet coupled to the Higgs field but the phenomenology of the special case of a radial mode of a field breaking a global symmetry, in a minimal perturbative model, is rarely discussed.} Furthermore, it is interesting to compare the extended EFT approach with a complete model which determines the correlations between different operators in the EFT approach.

The Minimal Linear $\sigma$ Model (ML$\sigma$M)~\cite{Feruglio:2016zvt} is in this respect a useful laboratory.
It is a rare example of an explicit non-supersymmetric perturbative renormalisable BSM with a rich structure.\footnote{The option of the Higgs boson as a pseudo-Goldstone is most often discussed in low energy models viewed as remnants of some kind of strong dynamics (Composite Higgs (CH)~\cite{Coleman:1969sm,Callan:1969sn,Kaplan:1983fs,Kaplan:1983sm,Banks:1984gj,Dugan:1984hq,Agashe:2004rs,Barbieri:2007bh,Gripaios:2009pe,Erdmenger:2020lvq,Erdmenger:2020flu}).}

The scalar potential of the ML$\sigma$M is based on the assumption of the underlying global $SO(5)$ symmetry, which is broken spontaneously $SO(5)\to SO(4)$ by a scalar quintuplet of $SO(5)$ and explicitly by a small number of soft terms (for the first through discussion of such a potential in the context of the fine-tuning and naturalness problems of the SM see Ref.~\cite{Barbieri:2007bh}).
One can then investigate the phenomenological effects of the mass eigenstate $\sigma$, mainly coinciding with the radial mode of the fifth component of the scalar quintuplet, which breaks spontaneously $SO(5)$. Since the presence of the scalar $\sigma$ in the spectrum restores perturbative unitarity, one can consistently discuss the hierarchy of masses $m_{\sigma}<M_{fermion}$ once the model is completed with a heavy vector-like fermion spectrum, which is the case in the ML$\sigma$M. In turn, one can investigate the EFT description of the model, with dim 5 and dim 6 operators present, after integrating out the heavy fermions, estimate the validity of the EFT approach and also compare it with the SMEFT.

With the question about the possibility of the electroweak baryogenesis in mind, the main goal of this paper is twofold. Firstly, we generalise the ML$\sigma$M Lagrangian, which was previously taken to be strictly CP conserving, to have arbitrary complex Lagrangian parameters, that is to have additional to the SM sources of CP violation. Secondly, after updating the phenomenological constraints on the parameters of the scalar potential we discuss the electroweak phase transition in the model. One important particular aspect of the potential is a flat direction in the limit of no explicit breaking of the $SO$(5) symmetry.

Concerning the first issue we focus on the predictions of the model on the CP violation in the effective SM fermion Yukawa couplings. It is well known that they are most strongly constrained by the electron Electric Dipole Moment (EDM). After introducing complex Lagrangian parameters, we compare the electron EDM bounds on the imaginary parts of the Wilson coefficients for different dim 5 and dim 6 operators treated as independent with those obtained when they are correlated by the model, pointing out the limitations of the effective description. With regard to the electroweak phase transition issue, we confirm the naive expectation that in the presence of the flat direction in the potential and only a weak explicit breaking of the $SO$(5) symmetry imposed by the phenomenological constraints, the Higgs potential does not have a tree level barrier. Furthermore the Coleman-Weinberg loop and thermal effects are by far too weak to give a strong enough electroweak baryogenesis.

The ML$\sigma$M has already been highly investigated in several different directions: its low-energy features have been studied in Ref.~\cite{Gavela:2016vte}, after projecting  it to the so-called Higgs Effective Field Theory Lagrangian~\cite{Feruglio:1992wf,Contino:2010mh,Alonso:2012px,Alonso:2012pz,Buchalla:2013rka,Brivio:2013pma,Brivio:2014pfa,Gavela:2014vra,Gavela:2014uta,Eboli:2016kko,Brivio:2016fzo,Merlo:2016prs,Alonso:2017tdy,Kozow:2019txg}; the possibility to solve the strong CP problem within the ML$\sigma$M has been analysed in Refs.~\cite{Brivio:2017sdm,Merlo:2017sun,Alonso-Gonzalez:2018vpc}; the phenomenology associated to the lightest exotic fermions has been illustrated in Ref.~\cite{Aguilar-Saavedra:2019ghg}. 

The structure of the paper can be read out in the table of contents.

\boldmath
\section{The Minimal Linear $\sigma$ Model}
\label{Sect.Model}
\unboldmath

The global symmetry group of the \MLsM is $SO(5)\times U(1)_X$, where the last Abelian factor ensures the correct hypercharge assignment for the SM fields. The gauge sector coincides with the one of the SM, while the scalar spectrum contains a real scalar field $\phi$ in the fundamental representation of $SO(5)$. The fermionic sector contains elementary fermions with the same quantum numbers as in the SM and exotic vector-like quarks in the trivial and fundamental representation of $SO(5)$.

The scalar field $\phi$ has five components which can be identified with the three would-be-longitudinal components of the SM gauge bosons $\pi_i$, $i=1,\,2,\,3$, the Higgs field $\nh$ and the 
additional scalar field $\ns$, singlet under the SM group:
\be
\phi=(\pi_1,\,\pi_2,\,\pi_3,\,\nh,\,\ns)^T
\xrightarrow{u.g.}
\phi=\left(0,\,0,\,0,\,\nh,\,\ns\right)^T\,,
\label{phiDefinition}
\ee
where the last expression holds in the unitary gauge and unbroken phase. 

The exotic fermions have $U(1)_X$ charges equal to $2/3$ or $-1/3$ and are associated to the up-type and down-type sectors, respectively. The proto-Yukawas are couplings between the $SO(5)$ quintuplet vector-like quarks (VLQs) $\psi$, the singlet VLQs $\chi$ and the scalar field $\phi$. The scalar field $\phi$ only couples directly to the exotic fermions, and in particular does not have direct couplings with the SM fermions. Therefore the SM Higgs couples to the SM fermions only through the mediation of the exotic fermions~\cite{Dugan:1984hq,Kaplan:1991dc,Contino:2004vy}. This structure is in line with one possible dynamical explanation of the hierarchical pattern of the SM Yukawa couplings, by mixing the SM chiral fermions with heavy exotic vector-like fermions with flavour-anarchical Yukawa couplings~\cite{Leurer:1992wg}. With the electroweak baryogenesis in mind and since the electron EDM constraints are the strongest for the third generation Yukawas~\footnote{See Ref.~\cite{Alonso-Gonzalez:2021jsa} for the discussion on how the electron EDM bounds apply to the other generations.}, we focus only on those.

The charged lepton sector can also be described by adding a copy of the down-quark Lagrangian with exotic fermions with $U(1)_X=-1$. For shortness, terms including charged leptons will not be shown explicitly and only comments will be added when necessary. 

Tab.~\ref{tab:SO5Transformations} summarises the scalar and exotic fermion fields appearing in the model with their transformation properties under $SO(5)\times U(1)_X$. 

\begin{table}[h!]
\centering{
\begin{tabular}{|c||c||c|c|c|c|}
\hline
&&&&&\\[-3mm]
		& $\phi$ 	& $\psi^{(2/3)}$ 	& $\chi^{(2/3)}$ 	& $\psi^{(-1/3)}$ 	& $\chi^{(-1/3)}$ \\[1mm]
\hline
&&&&&\\[-3mm]
$SO(5)$ 	& $5$	& $5$ 		& $1$ 		& $5$ 		& $1$ \\
$U(1)_X$	& $0$ 	& $+2/3$ 		& $+2/3$ 		& $-1/3$ 		& $-1/3$ \\[1mm]
\hline
\end{tabular}
\caption{\em Transformation properties of the scalar and exotic fields under $SO(5)\times U(1)_X$. 
The superscripts $(2/3)$ and $(-1/3)$ on the fermionic fields refer to the top and bottom quark sectors.}
\label{tab:SO5Transformations}
}
\end{table}

The \MLsM is a perturbative renormalisable model but it does not address the question of the stability of the $SO(5)$ breaking scale. It can then be viewed as a ``low-energy" renormalisable effective theory, valid below some naturalness cut-off scale where new physics cuts off the quadratic divergence destabilising that scale. 

\subsection{The Scalar Sector}
\label{SubSect.Scalar}

The spontaneous breaking of $SO(5)$ to $SO(4)$ and of the EW symmetries is described by the scalar potential $V(\phi)$~\cite{Barbieri:2007bh,Feruglio:2016zvt}:
\be
V(\phi)=\lambda\left(\phi^T\phi-f^2\right)^2+\alpha\,f^3\,\ns-\beta\,f^2\,\nh^2\,,
\label{ScalarPotential}
\ee
where $f$ is the scale at which the $SO(5)$ breaking takes place for $\alpha=\beta=0$. The last two terms, proportional to $\alpha$ and $\beta$, are the soft terms which break the $SO(5)$ symmetry explicitly. Those terms provide the potential for the Goldstone bosons present in the spectrum after the $SO(5)\rightarrow SO(4)$ breaking and 
are in principle calculable as quantum Coleman-Weinberg corrections in terms of the Lagrangian parameters of the gauge and fermion sector, that also break $SO(5)$ explicitly. They are sufficient to absorb the dominant one-loop Coleman-Weinberg contributions (see Ref.~\cite{Merlo:2017sun} for a different treatment). Since our fermion sector is not complete (we do not consider the first two generations), in this section we take them as free parameters and determine their range consistent with the available phenomenological constraints. After introducing our fermion sector, we return to the question of the compatibility of their values with the Coleman-Weinberg calculation.

Another interesting aspect of the potential in Eq.~\eqref{ScalarPotential} is the compatibility of the phenomenologically
acceptable range of the parameters with the naturalness and fine-tuning criteria discussed in detail in Ref.~\cite{Barbieri:2007bh}. Since those concepts are only of a qualitative nature, guided by the discussion in Ref.~\cite{Barbieri:2007bh}, we consider the scale $\Lambda_{nat}\sim{\cal O}(10)f$ as an acceptable upper bound for the effective ML$\sigma$M and the onset of new physics. This implies in particular the upper bound $\Lambda_{nat}$ for the physical masses of the exotic fermions introduced in the next section. Similarly, guided by Ref.~\cite{Barbieri:2007bh}, one expects the values $\lambda, \alpha,\beta\sim{\cal O}(1)$ to be acceptable from the fine-tuning perspective. 

Equipped with those qualitative considerations, we now proceed to the phenomenological analysis of the potential in Eq.~\eqref{ScalarPotential}.

With the symmetry breakings, $\nh$ and $\ns$ acquire non-vanishing VEVs:
\be
\nh=\hat{\nh}+v_\nh\,,\qquad\qquad
\ns=\hat{\ns}+v_\ns\,,
\label{VEVS}
\ee
where the normalisation has been chosen to match Eq.~(\ref{phiDefinition}). While these two VEVs are undefined if $\alpha=0=\beta$, their general expression turns out to be 
\be
\label{VEVs_hs}
v_\ns^2=f^2\dfrac{\alpha^2}{4\beta^2}\,,\qquad\qquad
v_\nh^2=f^2\left(1-\dfrac{\alpha^2}{4\beta^2}+\dfrac{\beta}{2\lambda}\right)\,,
\ee
with 
\be
v_\nh^2+v_\ns^2=f^2\left(1+\dfrac{\beta}{2\lambda}\right)\,.
\ee
We see that the effective scale $v_\ns$ of the $SO(5)$ breaking depends also on the soft breaking parameters $\alpha$ and $\beta$. 

The $SO(5)$ breaking requires that $f^2>0$, while imposing that the Higgs arises as a GB leads to $\left| v_\nh \right|<\left| v_\ns \right|$: it follows that 
\be
2\beta^2\left(1+\frac{\beta}{2\lambda}\right)<\alpha^2<4\beta^2\left(1+\frac{\beta}{2\lambda}\right).
\label{TheoContraintsab}
\ee

Once the symmetries are broken, a non diagonal $2\times2$ mass matrix can be read out from Eq.~\eqref{ScalarPotential}. The mass eigenstates $h$ and $\sigma$, after diagonalising this mass matrix, can be defined as
\be
h=\hat\nh\,\cos\gamma-\hat\ns\sin\gamma\,,\quad
\s=\hat\nh\,\sin\gamma+\hat\ns\cos\gamma\,,
\label{PhysicalScalars}
\ee
where the mixing angle is given by
\be
\tan 2\gamma=\dfrac{4\,v_\nh\,v_\ns}{3v_\ns^2-v_\nh^2-f^2}\,.
\ee
The mass eigenvalues are complicated expressions of the parameters appearing in the scalar potential and of the scalar VEVs. 

\begin{figure}[h!]
\centering
\includegraphics[width=0.49\textwidth]{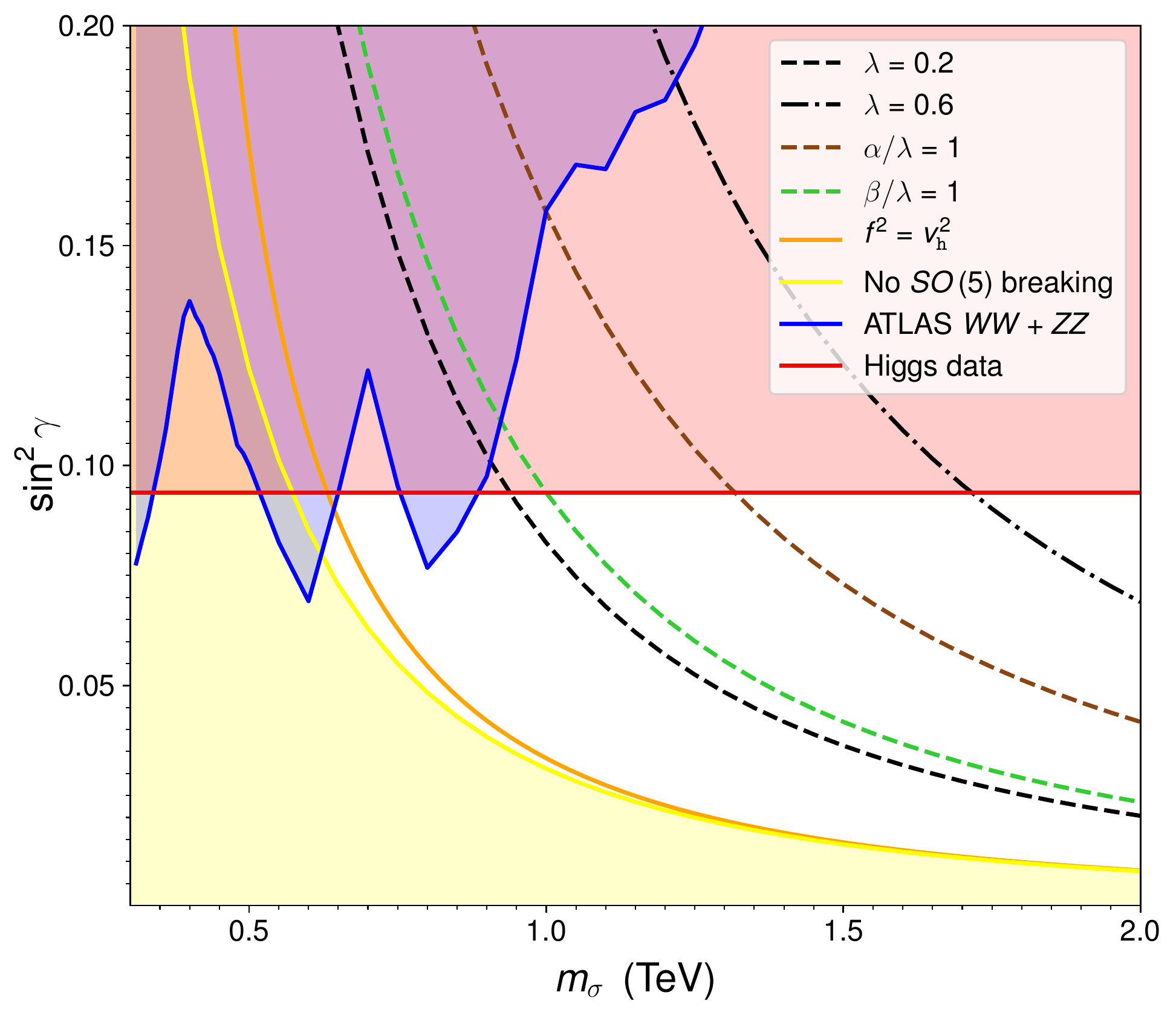} 
\caption{\em Constraints on the mass ($m_\sigma$) and mixing angle ($\sin^2 \gamma$) of the new singlet scalar $\sigma$ (see text for details).}
\label{fig:sg2_VS_ms}
\end{figure}

It is useful to express the different parameters of the scalar potential in terms of the Fermi constant $G_\text{F}$, the scalar masses, the mixing angle $\gamma$:
\be
\begin{aligned}
&v^2_\nh=\dfrac{1}{\sqrt2\,G_\text{F}}&\qquad
&v_\ns=\dfrac{v_\nh\sin(2\gamma)(m_\s^2-m_h^2)}{m_\s^2+m_h^2-(m_\s^2-m_h^2)\cos(2\gamma)}\\
&\lambda=\dfrac{\sin^2\gamma\,m_\s^2}{8v_\nh^2}\left(1+\cot^2\gamma\,\dfrac{m_h^2}{m_\s^2}\right)&\qquad
&\dfrac{\beta}{4\lambda}=\dfrac{m_\s^2m_h^2}{\sin^2\gamma\,m_\s^4+\cos^2\gamma\,m_h^4-2m_h^2m_\s^2}\\
&\dfrac{\alpha^2}{4\beta^2}=\dfrac{\sin^2(2\gamma)(m_\s^2-m_h^2)^2}{4\left(\sin^2\gamma\,m_\s^4+\cos^2\gamma\,m_h^4-2m_h^2m_\s^2\right)}&\qquad
&f^2=\dfrac{v_\nh^2\left(\sin^2\gamma\,m_\s^4+\cos^2\gamma\,m_h^4-2m_h^2m_\s^2\right)}{\left(\sin^2\gamma\,m_\s^2+\cos^2\gamma\,m_h^2\right)^2}\,.
\end{aligned}
\ee
As $G_\text{F}$ and $m_h$ are fixed experimentally, the two free parameters are the mass of the singlet $\s$ and of its mixing angle with the Higgs boson. Fig.~\ref{fig:sg2_VS_ms} collects the present bounds on the parameter space ($m_\s , \sin^2\gamma$).

The red horizontal line corresponds to the upper bound on $\sin^2 \gamma$ that universally suppresses
the couplings of $h$ to the SM particles with respect to their SM values and it comes from the 125 GeV Higgs signal strengths at the LHC by ATLAS and CMS. Very recent ATLAS results from a $\sqrt{s} = 13$ TeV analysis of Higgs signal strengths with $80$ fb$^{-1}$ of integrated luminosity~\cite{ATLAS:2019slw} set the bound 
\be
\sin^2\gamma \lesssim 0.09
\label{sin2gammaBound}
\ee
at 95\% C.L. using a $\chi^2$ fit to the ATLAS data by assuming a universal suppression of Higgs couplings. The whole red shaded region is then excluded by this data. 

The blue region is the excluded area due to searches of heavy scalars decaying into SM gauge boson pairs. In order to derive this bound, it is useful to introduced an effective Higgs-singlet mixing angle $\sin^2 \gamma_\text{eff}$ 
as the ratio between the cross sections for the gluon fusion production of $\sigma$ in the \MLsM and for the gluon fusion production of a SM-like scalar $h_{m_\sigma}$, with mass $m_\sigma$,
\begin{equation}
\sin^2 \gamma_\text{eff} = \frac{\sigma(gg \to \sigma)_{\text{ML}\sigma\text{M}}}{\sigma(gg \to h_{m_\sigma})_\text{SM}}\,.
\label{ec:gammaeff}
\end{equation}
 In principle, such a bound should also take into account the VLQ loop contributions to $gg \to \sigma$, whose expressions can be found in Ref.~\cite{Feruglio:2016zvt}.
Neglecting the VLQ loop contributions, one simply has $\sin^2 \gamma_\text{eff} = \sin^2 \gamma$. 
Fig.~\ref{fig:sg2_VS_ms} shows the 95\% C.L. limits in $\sin^2 \gamma$ as an excluded blue area, after such an identification and using the latest $\sqrt{s} = 13$ TeV ATLAS search for scalar resonances in di-boson final states with 36 fb$^{-1}$ of integrated luminosity~\cite{Aaboud:2018bun}.
 The inclusion of the exotic fermion contributions is discussed in Ref.~\cite{Aguilar-Saavedra:2019ghg}.

Finally, Fig.~\ref{fig:sg2_VS_ms} highlights the impact of theoretical constraints on the \MLsM parameter space: the 
area under the yellow curve is ruled out when requiring spontaneous symmetry breaking of the \MLsM $SO(5)$ group. 
Note that $v_\nh^2/f^2 \equiv \xi = 1$ (with $\xi$ the non-linearity parameter typically introduced in CH models),
depicted by the orange line, and even regions where $v_\nh^2 > f^2$ are not excluded by experimental bounds. 

Fig.~\ref{fig:sg2_VS_ms} also shows some contours of constant values of
the $SO(5)$ coupling $\lambda$ and of the ratios $\alpha/\lambda =1$ and $\beta/\lambda =1$. We collect below the values of all the parameters of the scalar potential in Eq.~\eqref{ScalarPotential} for a few chosen points on the plot, for $\sin^2\gamma$ = 0.08:
\begin{equation}
\begin{aligned}
&m_\sigma= 0.7 \TeV\,\,\rightarrow\,\, &&\lambda = 0.11,\, \beta =0.83,\, \alpha = 3.36,\, f=293 \GeV,\, v_\ns= 590 \GeV\\
&m_\sigma= 1.5 \TeV\,\, \rightarrow\,\,&&\lambda = 0.40,\, \beta =0.17,\, \alpha = 0.35,\, f=733 \GeV,\, v_\ns= 768 \GeV \\
&m_\sigma= 2.0 \TeV\,\, \rightarrow\,\,&&\lambda = 0.69,\, \beta =0.15,\, \alpha = 0.30,\, f=791 \GeV,\, v_\ns= 795 \GeV\,.
\end{aligned}
\label{eq:scalar_param_values}
\end{equation}
We conclude that the scalar potential parameter range allowed by the phenomenological constraints is in very good agreement with the fine-tuning criteria discussed earlier.

\subsection{The Fermionic Sector}

\label{SubSect.Fermionic}

We are interested in the exotic fermion masses in the range $m_\sigma < m_{fermion} < \Lambda_{nat}$ where the upper bound comes from the expected cut-off for our effective perturbative model.
The part of the Lagrangian describing fermion interactions can be written as follows~\cite{Feruglio:2016zvt}
\begin{align}
\sL_f = & \,\,\ov{\nq}_L\,i\slashed{D}\,\nq_L + \ov{\nt}_R\,i\slashed{D}\,\nt_R + \ov{\nb}_R\,i\slashed{D}\,\nb_R +\notag  \\ 
 & +\ov{\psi}^{(2/3)}\left[i\slashed{D}-\mf\right]\psi^{(2/3)}+\ov{\chi}^{(2/3)}\left[i\slashed{D}-\mo\right]\chi^{(2/3)}+\notag \\
&+ \ov{\psi}^{(-1/3)}\left[i\slashed{D}-\mpf\right]\psi^{(-1/3)}+\ov{\chi}^{(-1/3)}\left[i\slashed{D}-\mpo\right]\chi^{(-1/3)} +\label{eq:lag_f} \\
 & -\Bigg[\yo\ov{\psi}^{(2/3)}_L\phi\chi^{(2/3)}_R+\yt\ov{\psi}^{(2/3)}_R\phi\chi^{(2/3)}_L+
\ypo\ov{\psi}^{(-1/3)}_L\phi\chi^{(-1/3)}_R+\ypt\ov{\psi}^{(-1/3)}_R\phi\chi^{(-1/3)}_L +\nn\\ 
 & +\lamo\left(\ov{\nq}_L\Delta^{(2/3)}_{2\times5}\psi^{(2/3)}_R\right)
+\lamt\ov{\psi}^{(2/3)}_L\left(\Delta^{(2/3)}_{5\times1}\nt_R\right)
+\lamth\ov{\chi}^{(2/3)}_L\nt_R +
\notag \\ 
 & 
+\lampo\left(\ov{\nq}_L\Delta^{(-1/3)}_{2\times5}\psi^{(-1/3)}_R\right)
+\lampt\ov{\psi}^{(-1/3)}_L\left(\Delta^{(-1/3)}_{5\times1}\nb_R\right)
+\lampth\ov{\chi}^{(-1/3)}_L\nb_R
+\hc\Bigg]\,.\nn 
\end{align}
where the $U(1)_X $ charge of the exotic fermion fields $\psi$ and $\chi$ has been made explicit throughout. The one associated to the charged lepton is very similar to the down-quark one with exotic leptons having $U(1)_X =-1$. The canonical kinetic terms for the SM quarks appear in the first line, where $\nq_L$ for the left-handed (LH) $SU(2)_L$-doublet, $\nt_R$ and $\nb_R$ for the RH $SU(2)_L$ singlets. The second and third lines describe the kinetic and mass terms for the exotic fermions. The forth line presents the proto-Yukawa interactions between the exotic quarks and the scalar quintuplet $\phi$. The last two lines contain the mixed terms describing the SM-exotic quark interactions: those terms proportional to $\Lambda^{(\prime)}_{1,2}$ explicitly break the $SO(5)$ symmetry and the spurions~\cite{DAmbrosio:2002vsn,Cirigliano:2005ck,Davidson:2006bd,Alonso:2011jd,Dinh:2017smk} $\Delta_{2\times5}$ and $\Delta_{5\times1}$ are introduced to formally restore the $SO(5)$ invariance:
\be
\begin{gathered}
\Delta^{(2/3)}_{2\times5}=\left(
\begin{array}{ccccc}
0& 0& 1& 0& 0 \\
0& 0& 0& 1& 0 \\
\end{array}\right)\,,\\
\Delta^{(-1/3)}_{2\times5}=\left(
\begin{array}{ccccc}
1& 0& 0& 0& 0 \\
0& 1& 0& 0& 0 \\
\end{array}\right)\,,\\
\Delta^{(2/3),(-1/3)}_{5\times1}=\left(
\begin{array}{ccccc}
0& 0& 0& 0& 1 \\
\end{array}\right)^T\,.
\end{gathered}
\label{SpurionStructure}
\ee
According to the fermion partial compositeness paradigm no direct elementary fermion couplings to $\phi$ are allowed.

The mass parameters $M^{(\prime)}_{1,5}$ can be taken as real: indeed, if they were complex, then the sum between the mass terms and their hermitian conjugates would end up with the real part of the mass parameters. On the other hand, the rest of parameters should be taken as complex: $\text{y}^{(\prime)}_{1,2}=\left|\text{y}^{(\prime)}_{1,2}\right|e^{i\alpha^{(\prime)}_{1,2}}$ and $\Lambda^{(\prime)}_{1,2,3}=\left|\Lambda^{(\prime)}_{1,2,3}\right|e^{i\beta^{(\prime)}_{1,2,3}}$. Performing a field redefinition it is possible to reduce the number of phases to the minimum set of 4 phases: a possible choice is 
\be
\begin{aligned}
&\psi_{L,R}^{(2/3)}\Longrightarrow e^{-i\beta_1}\psi_{L,R}^{(2/3)}\,,\qquad\qquad
&&\psi_{L,R}^{(-1/3)}\Longrightarrow e^{-i\beta'_1}\psi_{L,R}^{(-1/3)}\,,\\
&\nt_R\Longrightarrow e^{-i(\beta_1+\beta_2)}\nt_R\,,\qquad\qquad
&&\nb_R\Longrightarrow e^{-i(\beta'_1+\beta'_2)}\nb_R\,,\\
&\chi_{L,R}^{(2/3)}\Longrightarrow e^{i(\beta_3-\beta_2-\beta_1)}\chi_{L,R}^{(2/3)}\,,\qquad\qquad
&&\chi_{L,R}^{(-1/3)}\Longrightarrow e^{i(\beta'_3-\beta'_2-\beta'_1)}\chi_{L,R}^{(-1/3)}\,,
\end{aligned}
\ee
leaving only $\text{y}^{(\prime)}_{1,2}$ as complex parameters. Notice that the phase rotations on $\psi_{L}$ and $\psi_{R}$ ($\chi_{L}$ and $\chi_{R}$) are identical, so that the mass terms remain real. Among the SM fields, only the RH quarks are rotated and in particular $\nq_L$ is not redefined to avoid mixing of the phases of the up and down sectors and in order to not introduce any additional phase to the SM gauge boson couplings. 

It is useful to rewrite the expression in Eq.~\eqref{eq:lag_f} in terms of the $SU(2)_L$ components of the different fields:
\be
\begin{gathered}
\phi=(H^T,\,\widetilde{H}^T,\,\ns)^T\,, \\
\psi^{(2/3)}\sim\left(\nK,\,\nQ,\,\nT_5\right)^T\,,\qquad  \chi^{(2/3)}\sim \nT_1\,, \\
\psi^{(-1/3)}\sim\left(\nQ',\,\nK',\,\nB_5\right)^T, \qquad \chi^{(-1/3)}\sim \nB_1\,,
\label{PsiChiComponents}
\end{gathered}
\ee
where $H$ is the SM $SU(2)_L$ doublet, with $\widetilde{H}\equiv i\sigma_2H^\ast$, and $\nK^{(\prime)}$ and $\nQ^{(\prime)}$ are $SU(2)_L$ doublets and $\nT_{1,5}$ and $\nB_{1,5}$ are singlets. The whole set of charge assignments can be read in Tab.~\ref{tab:SMTransformations}, where the hypercharge is defined as 
 \begin{equation}
Y=\Sigma_R^{(3)}+X \,,
\end{equation}
with $X$ the $U(1)_X$ charge and $\Sigma_R^{(3)}$ the third component of the global $SU(2)_R$, which is part of the residual $SO(4)$ group after the breaking of $SO(5)$.

\begin{table*}[htb]
\centering 
\renewcommand{\arraystretch}{1.5}
\footnotesize
\begin{tabular}{|c|c|c|c|c|c|c| }
\hline
Charge/Field & $\nK$ & $\nQ$ & $\nT_{1,5}$ & $\nQ'$ & $\nK'$ & $\nB_{1,5}$ \\[0.5ex] 
\hline
$\Sigma^{(3)}_R$ & $+1/2$ & $-1/2$ & 0 & $+1/2$ & $-1/2$ & 0 \\
\hline
$SU(2)_L \times U(1)_Y$ & $(2,+7/6)$ & $(2,+1/6)$ & $(1,+2/3)$ & $(2,+1/6)$ & $(2,-5/6)$ & $(1,-1/3)$ \\
\hline 
$U(1)_X$ & $+2/3$ & $+2/3$ & $+2/3$ & $-1/3$ & $-1/3$ & $-1/3$ \\
\hline
$U(1)_{EM}$ & 
$\begin{matrix} \nK^u = +5/3 \\ \nK^d = +2/3 \end{matrix}$ & 
$\begin{matrix} \nQ^u = +2/3 \\ \nQ^d = -1/3 \end{matrix}$ & 
$+2/3$ & 
$\begin{matrix} \nQ'^u = +2/3 \\ \nQ'^d = -1/3 \end{matrix}$ & 
$\begin{matrix} \nK'^u = -1/3 \\  \nK'^d = -4/3 \end{matrix}$ & 
$-1/3$ \\
\hline 
\end{tabular}
\caption{\em Decompositions of the exotic quarks and their transformations under the SM group.}
\label{tab:SMTransformations}
\end{table*}

By using Eqs.~\eqref{SpurionStructure} and \eqref{PsiChiComponents}, Eq.~\eqref{eq:lag_f} acquires the following form
\be
\begin{aligned}
\hspace{-7mm}
\sL_f=&\phantom{+}\ov{\nq}_L\,i\slashed{D}\,\nq_L+\ov{\nt}_R\,i\slashed{D}\,\nt_R+\ov{\nb}_R\,i\slashed{D}\,\nb_R+\\
&+\ov{\nK}\left[i\slashed{D}-\mf\right]\nK+\ov{\nQ}\left[i\slashed{D}-\mf\right]\nQ
+\ov{\nT}_5\left[i\slashed{D}-\mf\right]\nT_5+\ov{\nT}_1\left[i\slashed{D}-\mo\right]\nT_1+\\ 
&+\ov{\nQ'}\left[i\slashed{D}-\mpf\right]\nQ'+\ov{\nK'}\left[i\slashed{D}-\mpf\right]\nK'
+\ov{\nB}_5\left[i\slashed{D}-\mpf\right]\nB_5+\ov{\nB}_1\left[i\slashed{D}-\mpo\right]\nB_1+\\ 
&-\Big[\yo\left(\ov{\nK}_L\,H\,\nT_{1,R}+\ov{\nQ}_L\,\tilde{H}\,\nT_{1,R}+\ov{\nT}_{5,L}\,\ns\, \nT_{1,R}\right)
+\yt\left(\ov{\nT}_{1,L}\,H^\dagger\, \nK_R+\ov{\nT}_{1,L}\,\tilde{H}^\dagger\, \nQ_R+\ov{\nT}_{1,L}\,\ns\, \nT_{5,R}\right)+\\ 
&\hspace{5mm}+\ypo\left(\ov{\nQ'}_L\,H\,\nB_{1,R}+\ov{\nK'}_L\,\tilde{H}\,\nB_{1,R}+\ov{\nB}_{5,L}\,\ns\, \nB_{1,R}\right)
+\ypt\left(\ov{\nB}_{1,L}\,H^\dagger\, \nQ'_R+\ov{\nB}_{1,L}\,\tilde{H}^\dagger\, \nK'_R+\ov{\nB}_{1,L}\,\ns\, \nB_{5,R}\right)+\\ 
&\hspace{5mm}+\lamo\ov{\nq}_L\nQ_R
+\lamt\ov{\nT}_{5,L}\nt_R
+\lamth\ov{\nT}_{1,L}\nt_R
+\lampo\ov{\nq}_L\nQ'_R
+\lampt\ov{\nB}_{5,L}\nb_R
+\lampth\ov{\nB}_{1,L}\nb_R
+\hc\Big]\,,
\end{aligned}
\label{eq:lag_f_SU2_components}
\ee
where the scalar fields $H$ and $\ns$ still denote the unshifted and unrotated fields defined in Eq.~(\ref{phiDefinition}). 
In order to match with the notation adopted in the previous section, notice that in the unitary gauge
\be
H\equiv
\left(
\begin{array}{c}
0 \\
\nh/\sqrt{2} \\
\end{array}\right)\,.
\ee

Fermion masses and mixings arise by diagonalising the whole fermion mass matrix that includes SM and exotic quarks. To this end, all fermions can be grouped together in a single vector
\begin{equation}
\Psi=\left(\nK^u,\,\cT,\,\cB,\,\nK'^d\right)^T\,,
\end{equation}
where the ordering of the components is based on their electric charges, $+5/3$, $+2/3$, $-1/3$, and $-4/3$, respectively. 
Notice that $\cT$ and $\cB$ list together all the states with the same electric charge, $+2/3$ and $-1/3$, respectively,
\be
\begin{aligned}
& \cT=\left(\nt,\, \nQ^u,\,\nK^d,\,\nT_5,\,\nT_1,\,\nQ'^{u}\right)^T\,, \\
& \cB=\left(\nb,\,\nQ'^d,\,\nK^u,\,\nB_5,\,\nB_1,\,\nQ^d\right)^T\,.
\end{aligned}
\ee
The whole fermion mass term can then be written in the interaction basis as 
\begin{equation}
\sL_\cM=-\ov{\Psi}_L\,\cM(v_\nh,v_\ns)\,\Psi_R,
\end{equation}
where the mass matrix $\cM(v_\nh,v_\ns)$ is a $14\times14$ block diagonal matrix,
\begin{equation}
\cM(v_\nh,v_\ns)=\diag\Big(\mf,\,\cM_\cT(v_\nh,v_\ns),\,\cM_\cB(v_\nh,v_\ns),\,\mpf\Big)\,,\label{eq:BigMassFermion}
\end{equation}
with
\begin{equation}
\cM_\cT(v_\nh,v_\ns)=
\left(
\begin{array}{cccccc}
0 	& \lamo 	& 0 		& 0 		& 0 			& \lampo\\
0 	& \mf 	& 0 		& 0 		& \yo\frac{v_\nh}{\sqrt2} 	& 0\\
0 	& 0		& \mf 	& 0		& \yo\frac{v_\nh}{\sqrt2} 	& 0\\
\lamt	& 0	& 	0	& \mf	& \yo v_\ns		& 0\\
\lamth	& \yt\frac{v_\nh}{\sqrt2}	& \yt\frac{v_\nh}{\sqrt2}	& \yt v_\ns 	& \mo	& 0\\
0	& 0		& 0		& 0		& 0			& \mpf
\end{array}
\right)\,,
\label{eq:FermionMassMatrixComponentsQuarkT}
\end{equation}
and similarly for $\cM_\cB(v_\nh,v_\ns)$, replacing the unprimed parameters with the primed ones and vice-versa. 

The matrix in Eq.~\eqref{eq:BigMassFermion} can be diagonalised through a bi-unitary transformation,
\begin{equation}
\widehat\Psi_{L,R}=U_{L,R}\Psi \quad\Longrightarrow\quad 
\widehat\cM=U_L\,\cM\, U^\dagger_R\,,
\end{equation}
where $\widehat\Psi_{L,R}$ stand for the mass eigenstates and $\widehat\cM$ for the diagonal matrix. The two unitary matrices can be written as block-diagonal structures
\be
U_{L,R}=\diag\left(1,U^\cT_{L,R},U^\cB_{L,R},1\right)\,,
\label{UnitaryMatricesULR}
\ee
where $U^\cT_{L,R}$ and $U^\cB_{L,R}$ diagonalise $\cM_\cT(v_\nh,v_\ns)$ and $\cM_\cB(v_\nh,v_\ns)$, respectively. Finally, the diagonalised mass matrix is given by 
\be
\widehat{\cM}=\diag\left(\mf,\widehat\cM_\cT,\widehat\cM_\cB,\mpf\right)
\ee
and the mass eigenstate fermion fields are defined as 
\be
\begin{aligned}
& \widehat\Psi=\left(K^u,\,\widehat\cT,\,\widehat\cB,\,K'^d\right)^T  \,,  \\
& \widehat\cT=\left( t,\,\sT,\,\sT_2,\,\sT_3,\,\sT_4,\,\sT_5 \right)^T  \,,  \\
& \widehat\cB=\left( b,\,\sB,\,\sB_2,\,\sB_3,\,\sB_4,\,\sB_5 \right)^T \,, 
\label{eq:phys_states}
\end{aligned}
\ee
with both charge $2/3$ and charge $-1/3$ mass eigenstates ordered by increasing masses, so that the lightest states correspond to the top and bottom quarks, respectively. Notice that the exotic charge states do not mix with the other fields and then $K^u\equiv \nK^u$ and $K^{\prime d}\equiv \nK^{\prime d}$. 

One can estimate the dependence of top partners masses on the fermionic Lagrangian parameters by analytically diagonalizing Eq.~\eqref{eq:FermionMassMatrixComponentsQuarkT} with $v_\nh=0$, which leads to these eigenvalues
\begin{equation}
\begin{split}
& \mf, \quad
\mf, \quad
\mpf, \\
& \dfrac{1}{2} \left(\mo + \mf -\sqrt{\mo^2-2\, \mo \mf + \mf^2 + 4\, v_\ns^2 \, \yo \yt} \right) , \\
& \dfrac{1}{2} \left(\mo + \mf +\sqrt{\mo^2-2\, \mo \mf + \mf^2 + 4\, v_\ns^2 \, \yo \yt} \right) ,
\end{split}
\label{eq:Approx_TP_masses}
\end{equation}
aside from the zero eigenvalue corresponding to the top quark mass in such limit. Similar results are obtained for bottom partners, replacing the unprimed parameters with the primed ones and vice-versa.

\subsection{The Low-Energy Effective Operators}

We collect in this section the Wilson coefficients for the operators obtained after the decoupling of the exotic fermions, calculated in terms of the original Lagrangian parameters.
The decoupling limit corresponds to considering $M\gg m_\s\gg v_\nh$, where $M$ represents the exotic fermion mass scales $M^{(\prime)}_i$ in Eq.~\eqref{eq:lag_f}. Generically, the GB scale $f$ is smaller than $M$ and therefore the additional expansion in the ratio $f/M\ll1$ will be also used.
This integration-out exercise has already been carried out in Ref.~\cite{Feruglio:2016zvt}, but without considering complex parameters: this section is devoted to reproducing the results in that paper, but taking into account the CPV sources. 

For energies $E<M$, the effective Lagrangian describing $d\leq 6$ operators containing SM quark interactions with gauge bosons, Higgs boson and $\ns$ can be written as
\be
\sL_\text{eff}=\ov q_L\, i \Ds\, q_L+\ov t_R\, i \Ds\, t_R+\ov b_R\, i \Ds\, b_R+\sum_{i,\, c_i\in\mathbb R}c_i\cO_i+\left(\sum_{j,\, c_j\in\mathbb C}c_j\cO_j+\hc\right)\,.
\ee
For concreteness, the ``Warsaw basis''~\cite{Grzadkowski:2010es} will be used and charged lepton operators will not be shown explicitly.

The light fermion kinetic terms also receive corrections from the integration-out procedure that require wave function renormalisation in order to recover canonically normalised kinetic terms,
\be
\begin{aligned}
&q_L\rightarrow \cZ^{-1/2}_{q_L}\,q_L\qquad\text{with}\qquad &&\cZ_{q_L}=1+\dfrac{\Lambda_1^2}{M_5^2}+\dfrac{\Lambda^{\prime2}_1}{M^{\prime2}_5}\\
&t_R\rightarrow \cZ^{-1/2}_{t_R}\,t_R\qquad\text{with}\qquad &&\cZ_{t_R}=1+\dfrac{\Lambda_2^2}{M_5^2}+\dfrac{\Lambda^{2}_3}{M^{2}_1}\\
&b_R\rightarrow \cZ^{-1/2}_{b_R}\,b_R\qquad\text{with}\qquad &&\cZ_{b_R}=1+\dfrac{\Lambda_2^{\prime2}}{M_5^{\prime2}}+\dfrac{\Lambda^{\prime2}_3}{M^{\prime2}_1}\,.
\end{aligned}
\label{WaveFunctionRenorm}
\ee

\begin{table}[h!]
\hspace{-1cm}
\renewcommand{\arraystretch}{2.75}
\begin{tabular}{|c | c| c| c|}
\hline
$d$ & Operator & $c_i$ & Leading Order in $f/M$ \\[0.5ex]
\hline \hline
\multirow{2}{*}{4}
&$\bar q_L~\wH~t_R$ & $-Y_t$ & $-  \left(\frac{\yo \Lambda_1\Lambda_3}{M_1 M_5}\right) {\mathcal Z}^{-1/2}_{q_L} {\mathcal Z}^{-1/2}_{t_R}$ \\
\cline{2-4}
&$\bar q_L~H~b_R$ & $-Y_b$ & $-\left(\frac{\yo' \Lambda'_1\Lambda'_3}{M'_1 M'_5}\right) {\mathcal Z}^{-1/2}_{q_L} {\mathcal Z}^{-1/2}_{b_R}$ \\
\hline \hline
\multirow{2}{*}{5}
&$\ns\,(\bar q_L \wH t_R)$ &
$c^{t}_{\ns 1}$ & $\frac{Y_t}{M_5}\left[\yt^\ast\frac{\Lambda_2}{\Lambda_3}- \left(\yo\frac{\Lambda_2\Lambda_3}{M_1 M_5}
+ \yt^\ast\frac{\Lambda_2\Lambda_3}{M^2_1}\right) \mathcal Z^{-1}_{t_R}\right] $\\
\cline{2-4}
&$\ns\,(\bar q_L H b_R)$ &
$c^{b}_{\ns 1}$ & $\frac{Y_b}{M'_5}\left[\yt^{\prime\ast}\frac{\Lambda'_2}{\Lambda'_3}-  \left(\yo'\frac{\Lambda'_2\Lambda'_3}{M'_1 M'_5}
+ \yt^{\prime\ast}\frac{\Lambda'_2\Lambda'_3}{{M'}^2_1} \right) \mathcal Z^{-1}_{b_R}\right]$\\
\hline\hline
\multirow{6}{*}{6}
&$\ns^2\,(\bar q_L \wH t_R)$ &
$c^{t}_{\ns 2}$ &\makecell{$-\frac{Y_t}{M_1 M_5}\Big\{\yo \yt^\ast - \left[
\yo \yt^\ast\left( 2\frac{\Lambda_2^2}{M_5^2} + \frac{\Lambda_3^2}{M_1^2} \right) +
\frac{\left(|\yt|^2+2\yt^{\ast2}\right) \Lambda_2^2 + |y_1|^2\Lambda_3^2}{2M_1 M_5} \right] \mathcal Z^{-1}_{t_R}+$ \\
\hspace{2cm} $+2\frac{\Lambda_2^2\Lambda_3^2}{M_1 M_5}
		\left(\frac{y_1\Re[\yo]}{M_5^2}+\frac{y_1\Re[\yt]+ \yt^\ast\Re[\yo]}{M_5 M_1}
	+\frac{\yt^\ast\Re[\yt]}{M_1^2}\right)\mathcal Z^{-2}_{t_R}\Big\} $}\\
\cline{2-4}
&$\ns^2\,(\bar q_L H b_R)$ &
$c^{b}_{\ns 2}$ &\makecell{$-\frac{Y_b}{M'_1 M'_5}\Big\{\yo' \yt^{\prime\ast}  -\left[
\yo' \yt^{\prime\ast}\left( 2\frac{\Lambda_2^{\prime2}}{M_5^{\prime2}} + \frac{\Lambda_3^{\prime2}}{M_1^{\prime2}} \right) +
\frac{\left(|\yt'|^2+2\yt^{\prime\ast2}\right) \Lambda_2^{\prime2} + |\yo'|^2\Lambda_3^{\prime2}}{2M'_1 M'_5} \right] \mathcal Z^{-1}_{b_R}$+ \\
\hspace{2cm} $ +2\frac{\Lambda_2^{\prime2}\Lambda_3^{\prime2}}{M'_1 M'_5}
		\left({\frac{\yo'\Re[\yo']}{M_5^{\prime2}}+\frac{\yo'\Re[\yt']+\yt^{\prime\ast}\Re[\yo']}{M'_5 M'_1}
	+\frac{\yt^{\prime\ast}\Re[\yt']}{M_1^{\prime2}}}\right)\mathcal Z^{-2}_{b_R}\Big\} $}\\
\cline{2-4}
&$\,|H|^2\,(\bar q_L \wH t_R)$ &
$c^{t}_{H2}$ &\makecell{ $-\frac{Y_t}{M_1 M_5}\Big[2\yo \yt^\ast
-\left(2\yo\yt^\ast \frac{\Lambda_3^2}{M_1^2}
+|\yo|^2\frac{\Lambda_3^2}{M_1 M_5}\right)\mathcal Z^{-1}_{t_R}$ \\
\hspace{3.5cm} $-\left(\yo\yt^\ast\frac{\Lambda_1^2}{M_5^2}
+\frac{|\yo|^2}{2}\frac{\Lambda_1^2}{M_1 M_5}\right)\mathcal Z^{-1}_{q_L}\Big]$}\\
\cline{2-4}
&$\,|H|^2\,(\bar q_L H b_R)$ &
$c^{b}_{H2}$ &\makecell{$-\frac{Y_b}{M'_1 M'_5}\Big[2\yo' \yt^{\prime\ast}
-\left(2\yo'\yt^{\prime\ast}\frac{\Lambda_3^{\prime2}}{M_1^{\prime2}}
+|\yo'|^2\frac{\Lambda_3^{\prime2}}{M'_1 M'_5}\right)\mathcal Z^{-1}_{b_R}$ \\
\hspace{3.5cm} $-\left(\yo'\yt'\frac{\Lambda_1^{\prime2}}{M_5^{\prime2}}
+\frac{|\yo'|^2}{2}\frac{\Lambda_1^{\prime2}}{M'_1 M'_5}\right)\mathcal Z^{-1}_{q_L}\Big]$}\\
\cline{2-4}
&$(H^\dagger i\overleftrightarrow {D_\mu}\,H)(\bar q_L \gamma^\mu q_L)$ &
$c^{(1)}_L$ & $\frac{1}{4}\left(\frac{|\yo|^2\Lambda_1^2}{M_1^2 M_5^2}-\frac{|\yo'|^2\Lambda_1^{\prime2}}{M_1^{\prime2} M_5^{\prime2}}\right){\mathcal Z}^{-1}_{q_L}$\\
\cline{2-4}
&$ (H^\dagger i\overleftrightarrow {D^i_\mu}\,H)(\bar q_L \tau^i \gamma^\mu q_L) $ &
$c^{(3)}_L$ & $-\frac{1}{4}\left(\frac{|\yo|^2\Lambda_1^2}{M_1^2 M_5^2}+\frac{|\yo'|^2\Lambda_1^{\prime2}}{M_1^{\prime2} M_5^{\prime2}}\right){\mathcal Z}^{-1}_{q_L}$\\
\hline
\end{tabular}
\caption{\em Leading order effective operators and their 
coefficients. The wave function renormalisation factors have been defined in Eq.~\eqref{WaveFunctionRenorm}.}
 \label{tabops}
\end{table}

The effective operators and their Wilson coefficients at leading order in $f/M$ are shown in Tab.~\ref{tabops}, where the following definitions are used:
\be
\begin{aligned}
(H^\dagger i\overleftrightarrow {D_\mu}\,H)&\equiv i\left(H^\dag\left(\overrightarrow{D_\mu}H\right)-\left(H^\dag\overleftarrow{D_\mu}\right)H\right)\,,\\
(H^\dagger i\overleftrightarrow {D^i_\mu}\,H)&\equiv i\left(H^\dag\tau^i\left(\overrightarrow{D_\mu}H\right)-\left(H^\dag\overleftarrow{D_\mu}\right)\tau^iH\right)\,.
\end{aligned}
\ee
Moreover, the symbol $\Re[x]$ refers to the real part of the parameter $x$. Notice that the coefficients of the $d=4$ operators also enter in the definition of the higher dimensional operator coefficients. 

The Yukawa couplings $Y_t$ and $Y_b$ are proportional to $\yo^{(\prime)}$ that are complex. The physical SM fermion masses receive also contributions from operators of higher dimensions once the scalar fields $H$ and $\ns$ develop their VEVs. The latter contributions are additionally suppressed by powers of $f/M$ and therefore can be safely neglected in the approximation considered here.

However, the complex phases entering $Y_t$ and $Y_b$ need to be reabsorbed. By redefining the fields $t_R$ and $b_R$ so that 
\be
t_R\rightarrow e^{-i\alpha_1}t_R\,,\qquad \qquad
b_R\rightarrow e^{-i\alpha'_1}b_R\,,
\label{lastRedef}
\ee
we get
\be
Y_t\rightarrow|Y_t|\equiv y_t\,,\qquad \qquad
Y_b\rightarrow|Y_b|\equiv y_b\,,
\ee
and the top and bottom masses at leading order in $f/M$ turn out to be
\begin{equation}
m_t  = \frac{|\yo|}{\sqrt2} \, \frac{\lamo\lamth}{\mo\mf} \, v_\nh \, {\mathcal Z}^{-1/2}_{q_L} {\mathcal Z}^{-1/2}_{t_R}\,,\qquad\qquad
m_b  = \frac{|\ypo|}{\sqrt2} \, \frac{\lamo'\lamth'}{\mo'\mf'}\, v_\nh \, {\mathcal Z}^{-1/2}_{q_L} {\mathcal Z}^{-1/2}_{b_R}\,.
\label{eq:leading_order_m_SM}
\end{equation}
The redefinitions in Eq.~\eqref{lastRedef} also affect the other operators in Tab.~\ref{tabops}, so that in the coefficients $c_{\ns1}^{t,b}$, $c_{\ns2}^{t,b}$ and $c_{H2}^{t,b}$, the couplings $Y_t$ and $Y_b$ should be replaced by their absolute value counterparts $y_t$ and $y_b$, respectively, and in the following they are always interpreted in this way. Since $\yo^{(\prime)}$ and $\yt^{(\prime)}$ present in those coefficients are complex, the new sources of CPV have not been washed out by the redefinition in Eq.~\eqref{lastRedef}.

It is worth stressing that in the considered ML$\sigma$M the effective suppression scale of the dim 5 operator and dim 6 operators are independent. In next section we investigate the following questions, always in the context of the electron EDM bounds on the magnitude of the acceptable CP violation:
\begin{enumerate}
\item What are the bounds on the imaginary parts of the different Wilson coefficients, if one assumes that each of them saturates by itself the experimental EDM bound?
\item What are the bounds on them when they are correlated by the model?
\item What is the role of the scalar $\s$ in the spectrum in each considered above cases, in particular how do the bounds in case 1. compare with the bound on the SMEFT operator $c^t_{H2}$?
\end{enumerate}

\boldmath
\section{The electron Electric Dipole Moment constraints}
\label{Sect.EDM}
\unboldmath

The strongest bounds on the beyond the SM sources of CP violation come from the fermion electric dipole moments (EDMs). The SM predictions give for them extremely small values and therefore any NP sources of CPV may show up at experiments as a clear signal of Beyond the SM (BSM) physics.
At present, the strongest experimental upper limit on the EDMs is the one for the electron. The result reported by the ACME II collaboration~\cite{Andreev:2018ayy} is:
\be
|d_e|<1.1\times 10^{-29}\text{ e cm} \qquad  \text{at }90\%\text{ C.L.}\,,
\label{BoundeEDM}
\ee
where $d_e$ is defined by the following effective Lagrangian
\be
\sL_\text{EDM}=-\dfrac{i\,d_e}{2}\bar{e}\,\sigma_{\mu\nu}\gamma_5\, e \,F^{\mu\nu}
\qquad\text{with}\qquad
\sigma_{\mu\nu}\equiv\dfrac{i}{2}\left[\gamma_\mu\gamma_\nu-\gamma_\nu\gamma_\mu\right]\,.
\ee

Although this limit is still far from the SM prediction which accidentally gives a very small value of $\approx 10^{-44}\text{ e cm}$~\cite{Zyla:2020zbs} (a recent reevaluation of this value have been done in Refs.~\cite{Yamaguchi:2020eub,Yamaguchi:2020dsy}), it constraints very strongly various BSM scenarios.

\subsection{General EFT analysis}

In this subsection we discuss the EDM constraints on the set of operators listed in Tab.~\ref{tabops}, which are generic for models with an additional electroweak singlet, assuming non-vanishing expectation value for the field $\ns$, $v_\ns>v_\nh$.

The electron EDM will receive NP contributions due to the imaginary parts of the Higgs and $\sigma$ boson couplings to the third generation SM fermions. The effective Lagrangian describing the fermion-scalar couplings can be parametrised as
\be
\sL_\text{eff}\supset-\dfrac{y_\psi}{\sqrt2}\left(\kappa^h_\psi\,\bar{\psi}\,\psi+i\,\tilde{\kappa}^h_\psi\,\bar\psi\,\gamma_5\,\psi\right)h + (h\rightarrow\sigma)
\qquad
\text{with}\qquad 
y_\psi\equiv \dfrac{\sqrt2\, m_\psi}{v_\nh}\,,
\label{EffectiveHPsiLag}
\ee
where $v_\nh=246\GeV$ is the EW symmetry breaking VEV of the Higgs field, and $m_\psi$ the mass of the SM fermion $\psi$. 

As long as only the CP violation effects in the Higgs boson Yukawa couplings are considered (e.g. in SMEFT), the electron EDM introduces a constraint on the three-dimensional space of the effective fermion parameters $\tilde\kappa^h_t, \tilde\kappa^h_b,\tilde\kappa^h_\tau$, with upper bounds on each of them (for a recent analysis see Ref.~\cite{Fuchs:2020uoc}). Particularly strong bounds are obtained for $ \tilde\kappa^h_t<{\cal O}(10^{-3})$ but the bounds for $ \tilde\kappa^h_\tau$ and $ \tilde\kappa^h_b$ are two orders of magnitude weaker. 
It is well known that the same effective parameters $\tilde\kappa_\psi$ can be responsible for the contribution of different third generation fermions to the baryon asymmetry in the universe in the electroweak phase transitions, \textit{provided} it is a strong enough first order transition (see Refs.~\cite{Fuchs:2020uoc,Sakharov:1967dj,Kuzmin:1985mm,Cohen:1993nk,Riotto:1999yt,Morrissey:2012db}) (which is not the case in the SM). The stringent bound on $\tilde\kappa^h_t$ excludes its dominant role in the electroweak baryogenesis but it can still be driven by
$ \tilde\kappa^h_\tau$ (see Refs.~\cite{deVries:2018tgs,Fuchs:2020uoc}). In this paper we are mainly focused on the comparison of the constraints on the complex parameters in the EFT approach with those in a renormalisable perturbative model and we shall do it taking the top Yukawa coupling as an example, as it gives the leading contribution to the electron EDM. The case of the $\tau$ Yukawa would exactly parallel our discussion, with appropriate rescalings.

\begin{figure}[h!]
\centering
\includegraphics[width=0.49\textwidth]{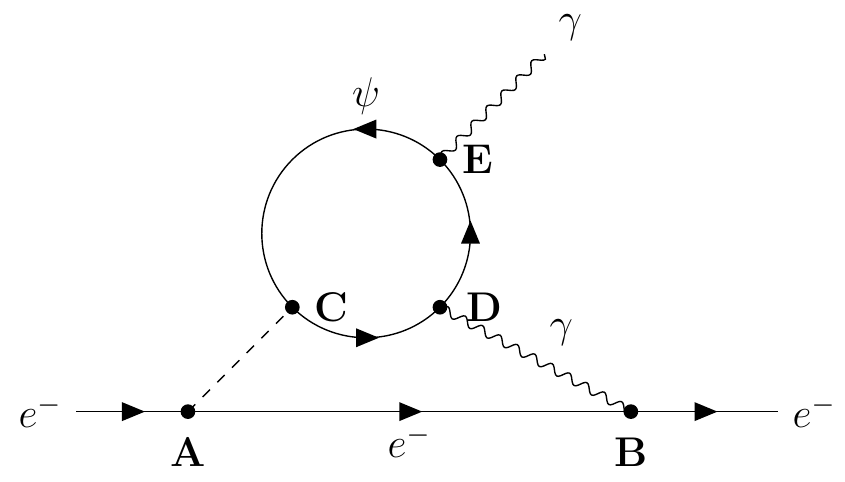}
\caption{\em Two-loop Barr-Zee diagram for the electron EDM.}
\label{fig:BZ}
\end{figure}

New contributions to the electron EDM arise through the so-called Barr-Zee diagram in Fig.~\ref{fig:BZ}. The different vertices are labelled with letters: we assume that the vertices``B'', ``D'' and ``E'' are purely SM (the photon couplings are left untouched in particular within the \MLsM). In a full three generation model, vertex ``A" may acquire an imaginary part due to new physics contributions, but it is expected to be suppressed accordingly to the fermion partial compositeness mechanism: the lightest fermions are mainly elementary, while the top, the bottom and tau are composite objects. The only vertex where new sources of CPV can play a role is ``C''. 

The explicit computation of the Barr-Zee diagram, allowing new CPV sources only in vertex ``C'', provides the following contribution to the electron EDM~\cite{Brod:2013cka}:
\begin{equation}
\dfrac{d_e}{e}=  -\sum_{\psi=t,b,\tau} 4 \, Q_e\,Q_\psi^2\,N_c\dfrac{\alpha_\text{em}}{(4\pi)^3}\sqrt2\,G_F\,m_e
\bigg[\kappa^h_e\,\tilde\kappa^h_\psi\,f_1(x_{\psi/h}) + \kappa^\sigma_e \,\tilde\kappa^\sigma_\psi\,f_1(x_{\psi/\sigma})\bigg]
\,,
\label{GenericEDM}
\end{equation}
where $Q_\psi$ is the $\psi$ electric charge, $N_c=3$ if $\psi$ is a quark and $N_c=1$ if $\psi$ is the $\tau$, $\alpha_\text{em}$ is the fine structure constant at the scale of the electron mass, $G_{\text F}$ the Fermi constant, and $f_1$ is a function of $x_{\psi/s}\equiv(m_\psi/m_s)^2$ defined by
\be\label{f1}
f_1(x)=\dfrac{2x}{\sqrt{1-4x}}\left[Li_2\left(1-\dfrac{1-\sqrt{1-4x}}{2x}\right)-Li_2\left(1-\dfrac{1+\sqrt{1-4x}}{2x}\right)\right]
\ee
being $Li_2$ the dilogarithm
\be
Li_2(z)=-\int_0^1\dfrac{\ln(1-zt)}{t}dt\,,\qquad z\in\mathbb{C}\,,\qquad t\in\mathbb{R}\,.
\ee
Taking the central experimental values for the fermion masses, $m_t=172.8\GeV$, $m_b=4.18\GeV$ and $m_\tau=1.78\GeV$, and the Higgs mass as $m_h=125.1\GeV$~\cite{Zyla:2020zbs}, and setting $m_\sigma = 1.5\TeV$ and $\sin^2\gamma=0.08$ as the benchmark values, the numerical results for the function $f_1$ computed for the top, bottom and $\tau$ fermions read: $f_1(x_{t/h})\simeq2.87$, $f_1(x_{b/h})\simeq0.055$ and $f_1(x_{\tau/h})\simeq 0.015$ when the Higgs boson is exchanged in the Barr-Zee diagram, while $f_1(x_{t/\sigma})\simeq0.30$, $f_1(x_{b/\sigma})\simeq0.0011$ and $f_1(x_{\tau/\sigma})\simeq 0.00026$ when it is the scalar singlet. These values have been obtained implementing in C++ the sub-routine Vegas \cite{Lepage:1980dq} from the Monte Carlo-based multi-dimensional Cuba library \cite{Hahn_2005}.

For a check of our calculation, we report the results for the upper bounds on the Higgs boson CPV effective couplings, assuming one contribution at the time, using the experimental bound in Eq.~\eqref{BoundeEDM}:
\begin{equation}
\tilde\kappa^h_t=0.0012\,,\qquad\qquad
\tilde\kappa^h_b=0.25\,,\qquad\qquad
\tilde\kappa^h_\tau=0.31 \,, \\
\label{eq:kappa_tilde_bounds}
\end{equation}
They agree very well with the values for $T^\psi_I$ obtained in Ref.~\cite{Fuchs:2020uoc} (see their Eq.~(4.5)), after the identification\footnote{Ref.~\cite{Fuchs:2020uoc} based their calculus on Ref.~\cite{Panico:2018hal}, where the formula in Eq.~(2.33) should be corrected including a factor $1/\sqrt{2}$ according to Ref.~\cite{Brod:2013cka}.}
\begin{equation}
T^\psi_I=\dfrac{\cos\gamma}{2\sqrt2} \, \tilde\kappa_\psi\, ,
\label{eq:Nir_et_al_coupling}
\end{equation}
since, due to the mixing between $h$ and $\sigma$, the SM Yukawa couplings for the electron end up suppressed as follows:
\begin{equation}
    \sL \supset -\frac{y_e}{\sqrt{2}} \nh \bar{e}_L e_R + \hc \supset -\frac{y_e}{\sqrt{2}}( h\bar{e}e \cos\gamma +  \sigma\bar{e}e \sin\gamma) \Rightarrow \begin{cases}
    \kappa_e^h = \cos\gamma \\ \kappa_e^\sigma = \sin\gamma \, .
    \end{cases}
    \label{kappasE}
\end{equation}\\

Our next task is to take the point of view of an EFT approach and to obtain the bounds on the imaginary parts of the Wilson coefficients, one by one, of the dimension 5 and 6 operators collected in Tab.~\ref{tabops}. Keeping only the top quark loop in the calculation, we get:
\begin{equation}
\begin{split}
\tilde\kappa^h_t  =  -(v_\ns \cos\gamma - v_\nh\sin \gamma) \, \Im[c^t_{\ns 1}]/y_t - v_\ns(v_\ns \cos\gamma - 2 \, v_\nh\sin\gamma) \, & \Im[c^t_{\ns 2}]/y_t +
\\
-\frac{3}{2} \, v_\nh^2\cos\gamma \, & \Im[c^t_{H2}]/y_t  \, ,
 \\[2.5pt]
\tilde\kappa^\sigma_t = -(v_\ns \sin\gamma + v_\nh\cos\gamma) \, \Im[c^t_{\ns 1}]/y_t - v_\ns(v_\ns \sin\gamma + 2 \, v_\nh\cos\gamma) \, & \Im[c^t_{\ns 2}]/y_t +
\\
- \frac{3}{2} \, v_\nh^2\sin\gamma \, & \Im[c^t_{H2}]/y_t \, ,
\end{split}
\label{kappasTop}
\end{equation}
where $\Im[x]$ is the imaginary part of $x$.

For illustrating the role played by the $h$ and $\sigma$ exchange, it is convenient to introduce the effective parameter $\tilde\kappa_t^{eff}$:
\begin{equation}
\tilde\kappa_t^{eff}=\kappa^h_e \, \tilde\kappa^h_t \, f_1(x_{t/h}) + \kappa^\sigma_e \, \tilde\kappa^\sigma_t \, f_1(x_{t/\sigma}) \, .
\label{kappaeff}
\end{equation}
For the benchmark point $m_\sigma =1.5 \TeV$ and $\sin^2\gamma$=0.08, if only one operator enters in the electron EDM, the $h$ ($\sigma$) contribution to $\tilde\kappa_t^{eff}$ is $98\%$ ($2\%$), $97\%$ ($3\%$) and $99\%$ ($1\%$) for $c^t_{\ns 1}, c^t_{\ns 2}$ and $c^t_{H2}$, respectively.

Although the contribution in Eq.~\eqref{kappasTop} of the different operators to the parameters $\tilde\kappa^h_t$ and $\tilde\kappa^\sigma_t$ is comparable, the contribution of the latter to the electron EDM is suppressed by the function $f_1$ and by $\kappa^\sigma_e=\sin\gamma$. From Eq.~\eqref{kappasTop} we also see that the coefficients  of $\Im[c^t_{\ns 2}]$ are larger than those of $\Im[c^t_{H2}]$ because $v_\ns>v_\nh$.  Therefore, the bounds on the  $\Im[c^t_{\ns 2}]$ should be stronger than the one on the $\Im[c^t_{H2}]$. Furthermore, we see that, because of the  mixing between $h$ and $\sigma$, the square of the coefficients of $\Im[c^t_{\ns 1}]$ are close to square of the coefficients of $\Im[c^t_{\ns 2}]$.

The bounds on $\Im[c^t_{\ns 1}]/y_t$ in units of $\text{TeV}^{-1}$ should  then be comparable to those on $\Im[c^t_{\ns 2}]/y_t$ in units of $\text{TeV}^{-2}$, leading to obvious implications for their interpretation in terms of the common suppression scale $\Lambda$ ($\Lambda^2$), respectively. This discussion is also very important for understanding the results of next section.

\begin{figure}[h!]
\centering
\minipage{0.5\textwidth}
\centering
\includegraphics[width=\linewidth]{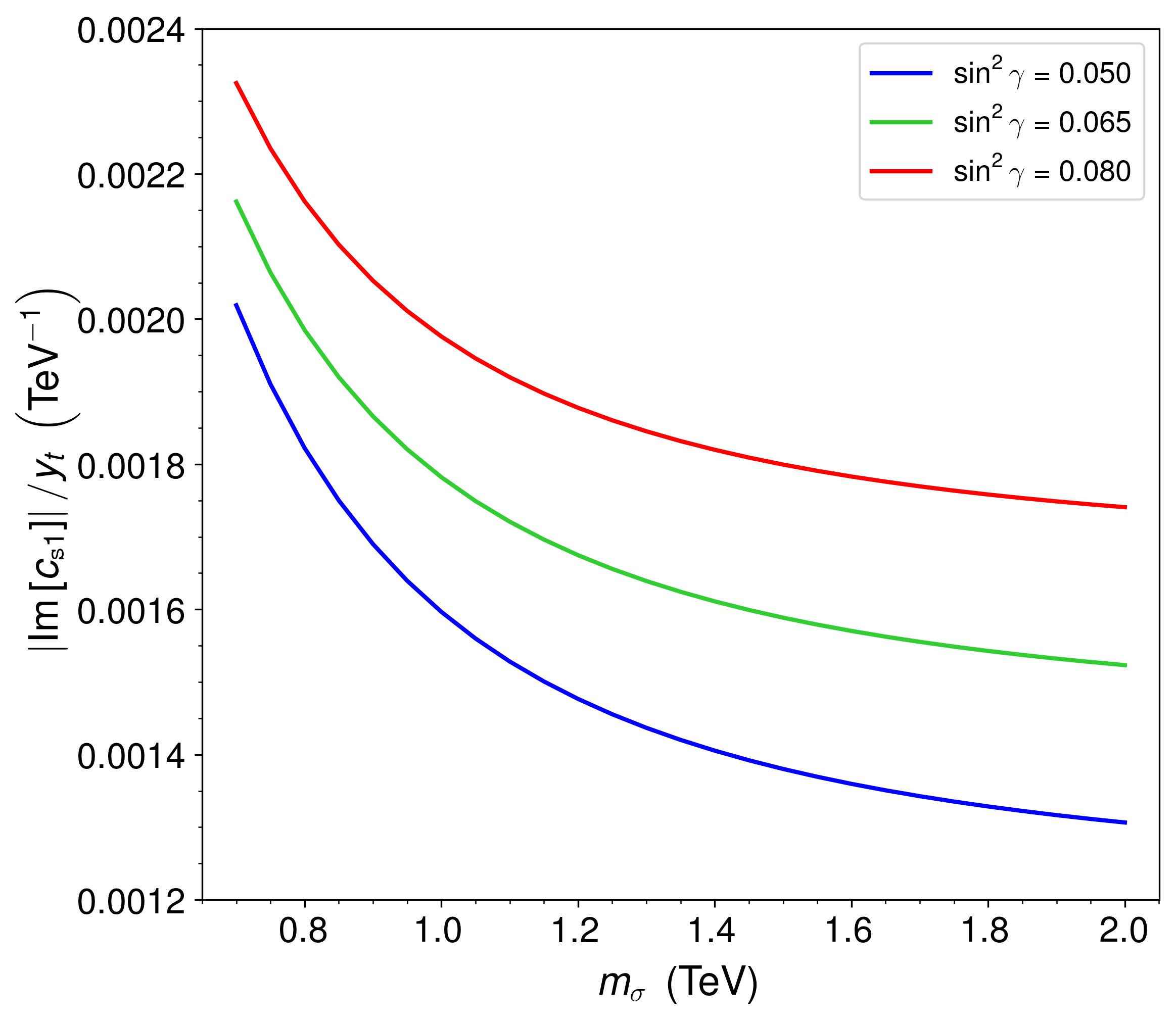}
\subcaption{}
\label{fig:EFT_ImC5s1}
\endminipage\hfill
\minipage{0.5\textwidth}
\centering
\includegraphics[width=\linewidth]{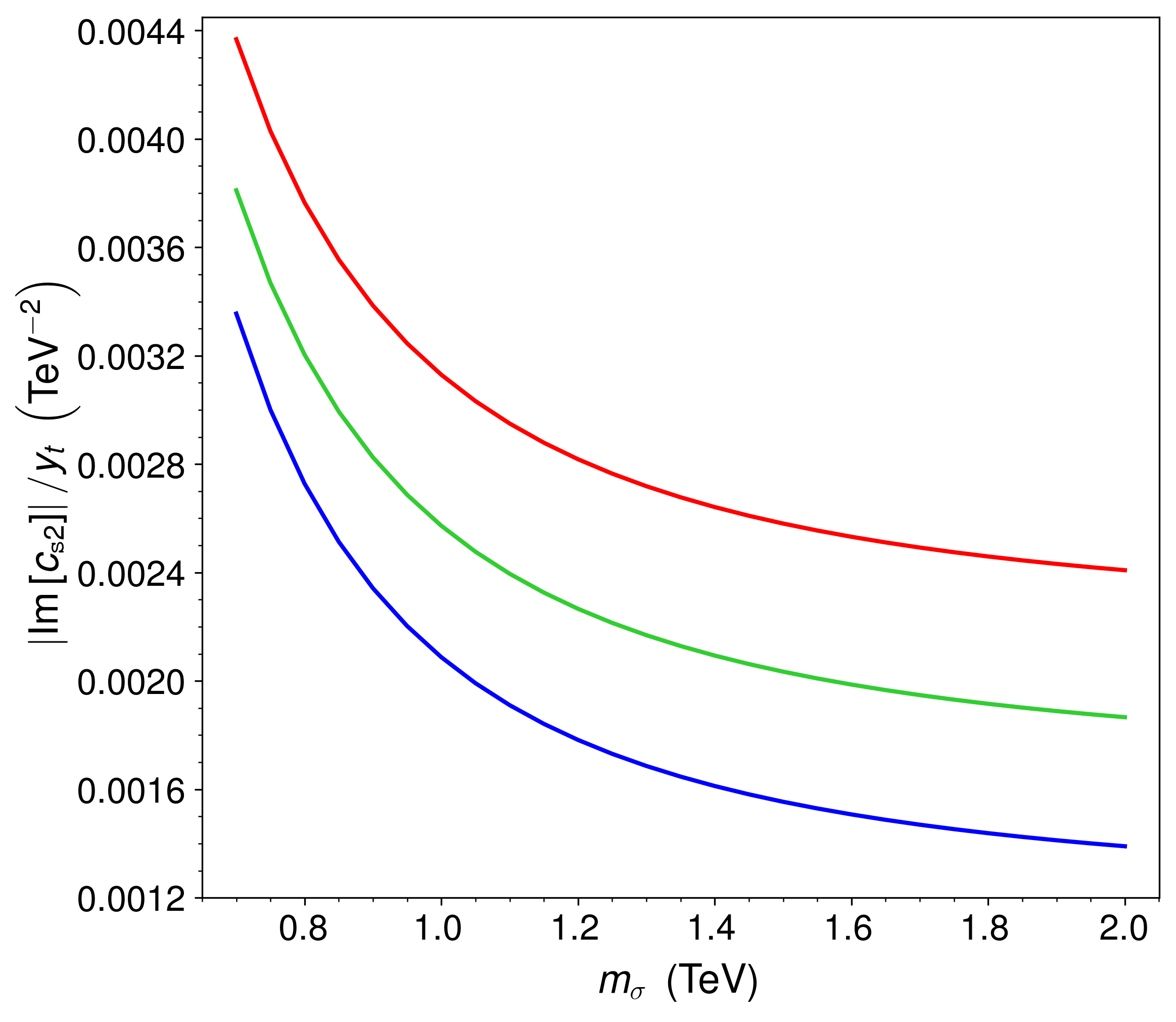}
\subcaption{}
\label{fig:EFT_ImC6s2}
\endminipage
\\
\minipage{0.5\textwidth}
\centering
\includegraphics[width=\linewidth]{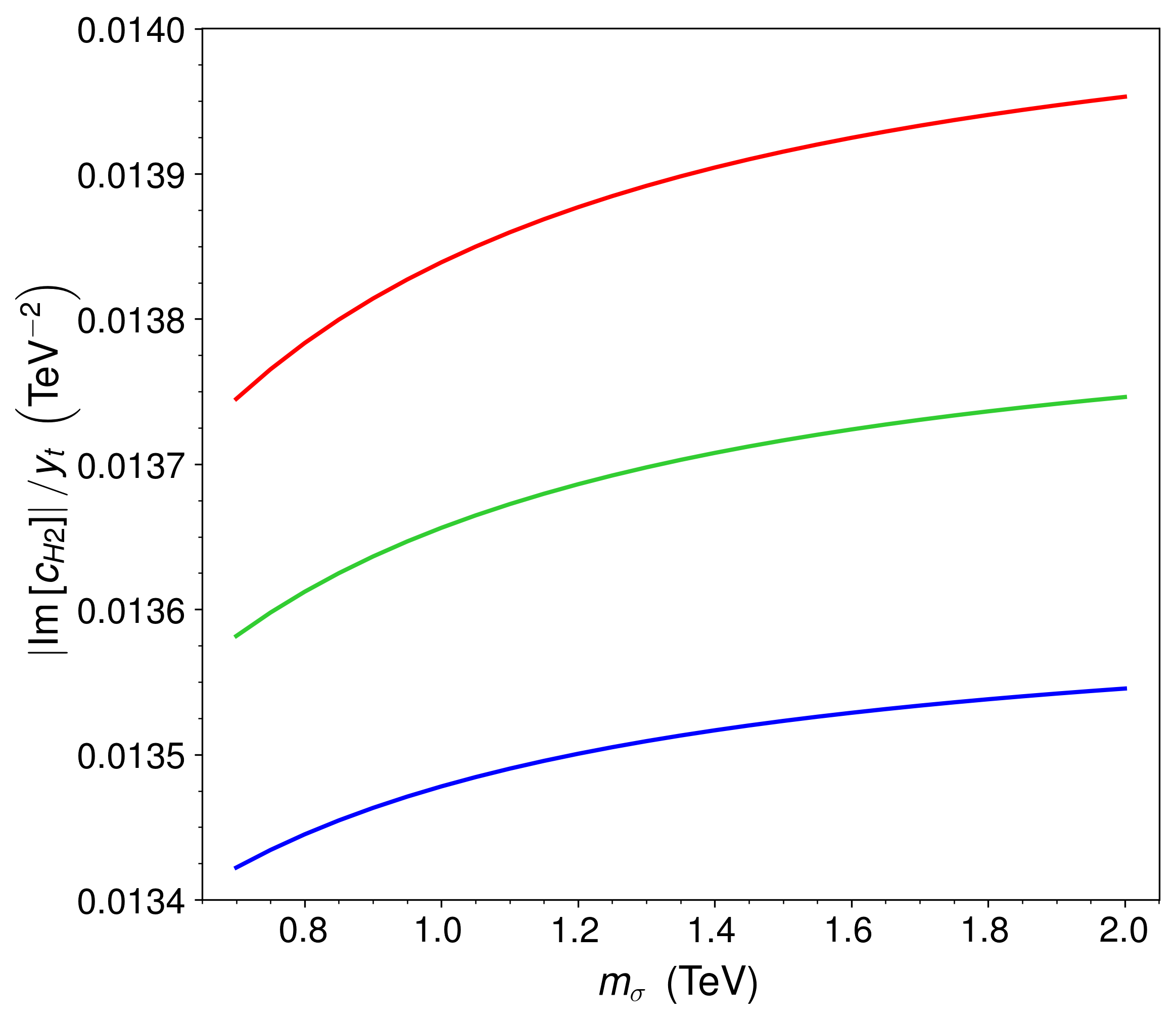}
\subcaption{}
\label{fig:EFT_ImC6H2}
\endminipage
\caption{{\it Bounds on the imaginary parts of the Wilson coefficients, $\absval{\Im[c^t_{\ns 1}]}/y_t$ (a), $\absval{\Im[c^t_{\ns 2}]}/y_t$ (b) and $\absval{\Im[c^t_{H 2}]}/y_t$ (c), assuming the contribution to the electron EDM of only one operator at a time.}}
\label{bound_wilsons}
\end{figure}

The bounds on the imaginary parts of the Wilson coefficients, assuming the contribution to the electron EDM of only one operator at a time, in the range of the parameters ($m_\sigma,\ \sin^2\gamma$) consistent with the Fig.~\ref{fig:sg2_VS_ms}, are shown in the Fig.~\ref{bound_wilsons}. They confirm the above described qualitative expectations. The $\Im[c^t_{\ns 1, 2}]/y_t$ have stronger suppression for larger $m_\sigma$ because the prefactor in Eq.~(\ref{kappasTop}) depends on $v_\ns$ that increases as $m_\sigma$ increases. On the other hand, $\Im[c^t_{H2}]/y_t$ has the opposite behaviour because in this case the dominant effect comes from the function $f_1(x_{t/\sigma})$ in Eq.~(\ref{kappaeff}) that decreases with increasing $m_\sigma$. 

Here are the bounds of the suppression scale $\Lambda$ following from the bounds on each of the Wilson coefficients, taking their phases as ${\cal O}(1)$: $\Lambda>(5.5\times 10^2, 20, 9) \TeV$, for $c^t_{\ns 1}$, $c^t_{\ns 2}$, $c^t_{H2}$, respectively, taking $m_\sigma=1.5 \TeV$ and $\sin^2_\gamma=0.08$. As expected, the strongest constrain on the new physics scale comes from the dim 5 operator. It is more than one order of magnitude stronger than the bounds from both $c^t_{\ns 2}$ and $c^t_{H2}$. The difference between the bounds coming from $c^t_{\ns 2}$ and $c^t_{H2}$ is due to the hierarchy
$v_\ns>v_\nh$.

In next section we will compare the obtained bounds with the results obtained using the complete ML${\sigma}$M Lagrangian.

\subsection{ML${\sigma}$M calculations}
\label{subsec:MLsM_calcs}

Our first question is: how the bounds on the imaginary parts of the Wilson coefficients change (compared to the results of the previous section) once we include the correlations between them which follow from the structure of the Lagrangian Eq.~\eqref{eq:lag_f}. This can be done by using the explicit expressions for the Wilson coefficients in terms of the Lagrangian parameters given in Tab.~\ref{tabops}.

The imaginary part of the Wilson coefficients in Eq.~(\ref{kappasTop}) are given by
\be
\begin{aligned}
\Im[c^t_{\ns 1}]/y_t =&
-\frac{1}{M_5}
\Bigg[
|\yo|\frac{\Lambda_2\Lambda_3}{M_1 M_5} {\mathcal Z}_{t_R}^{-1}  \sin\alpha_1 +
|\yt| \left(\frac{\Lambda_2}{\Lambda_3} - \frac{\Lambda_2\Lambda_3}{M_1^2}  {\mathcal Z}_{t_R}^{-1} \right) \sin\alpha_2
\Bigg]
\\
\Im[c^t_{\ns 2}]/y_t =&
-\frac{1}{M_1M_5}
\Bigg[
|\yo||\yt|
\left( 1 - \left[ 2 \frac{\Lambda_2^2}{M_5^2} + \frac{\Lambda_3^2}{M_1^2} \right]{\mathcal Z}_{t_R}^{-1} + 2\frac{\Lambda_2^2\Lambda_3^2}{M_1^2M_5^2}{\mathcal Z}_{t_R}^{-2} \right) \sin(\alpha_1 - \alpha_2) + \\
&\hspace{1cm}+ |\yo|^2
\frac{\Lambda_2^2\Lambda_3^2}{M_1M_5^3} {\mathcal Z}_{t_R}^{-2} \sin(2 \, \alpha_1) +
|\yt|^2
\left(\frac{\Lambda_2^2}{M_1M_5}{\mathcal Z}_{t_R}^{-1}  - \frac{\Lambda_2^2 \Lambda_3^2}{M_1^3 M_5} {\mathcal Z}_{t_R}^{-2} \right) \sin(2 \, \alpha_2)
\Bigg]
\\
\Im[c^t_{H2}]/y_t =& -\frac{2}{M_1M_5}
|\yo||\yt|
\left[ 1 - \frac{\Lambda_3^2}{M_1^2}{\mathcal Z}_{t_R}^{-1} - \frac{\Lambda_1^2}{2\,M_5^2}{\mathcal Z}_{q_L}^{-1} \right]
\sin(\alpha_1 - \alpha_2)
\end{aligned}
\label{eq:Im_WC_MLsM}
\ee

\begin{figure}[h!]
\centering
\minipage{0.5\textwidth}
\centering
\includegraphics[width=\linewidth]{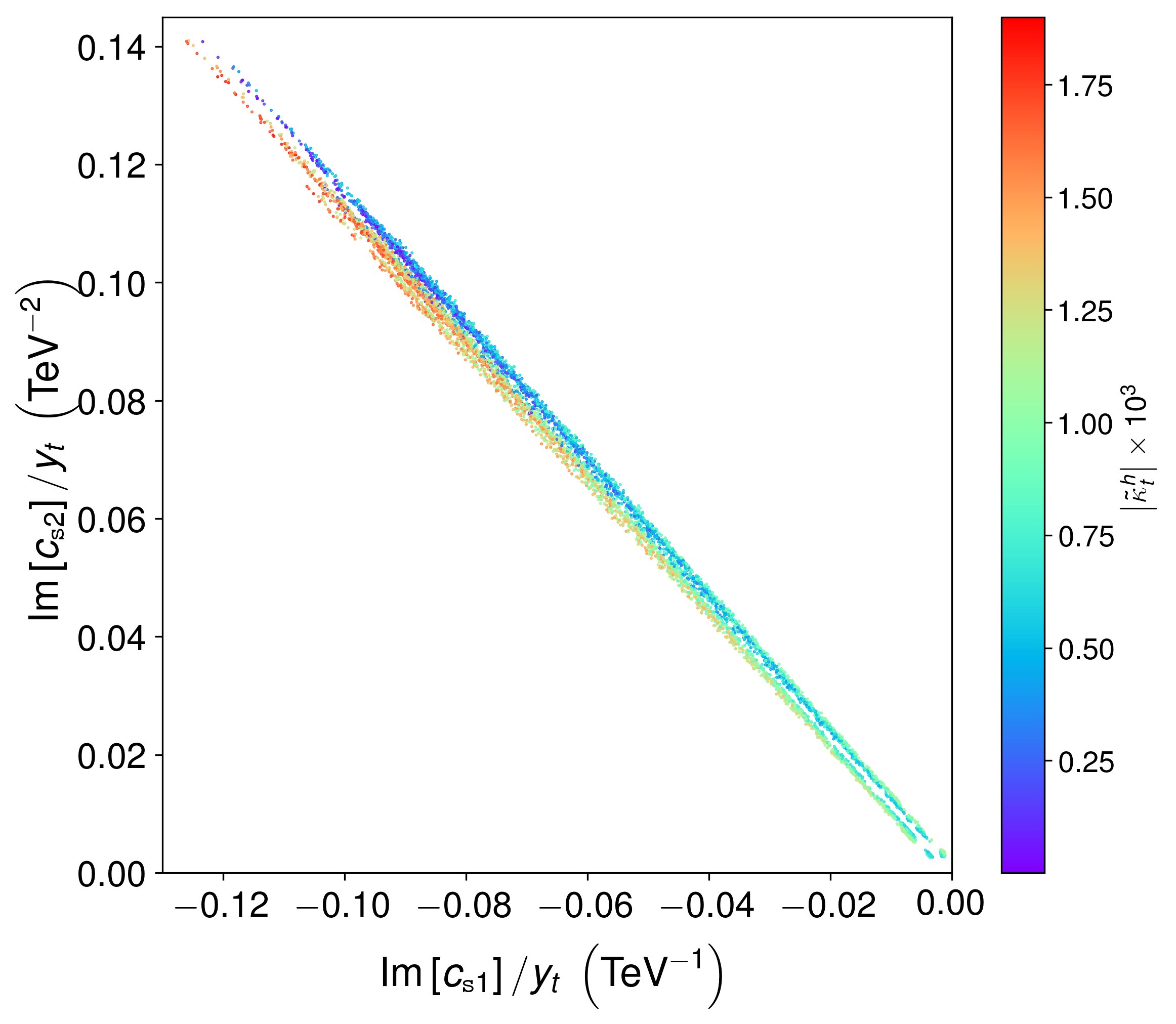}
\subcaption{}
\label{fig:ImC5s1_vs_ImC6s2_close_to_EDM_bound_with_kappa}
\endminipage\hfill
\minipage{0.5\textwidth}
\centering
\includegraphics[width=\linewidth]{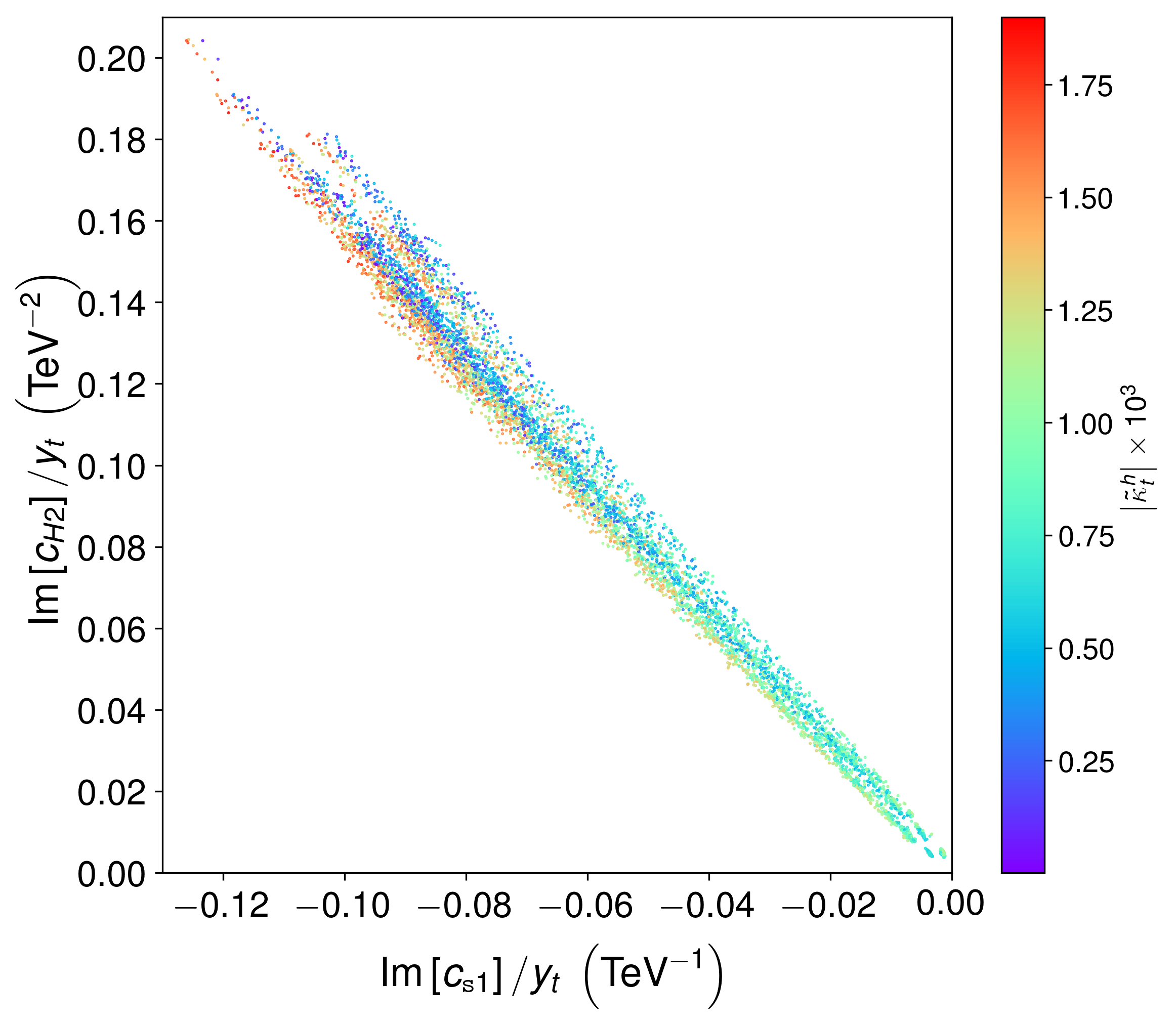}
\subcaption{}
\label{fig:ImC5s1_vs_ImC6H2_close_to_EDM_bound_with_kappa}
\endminipage
\\
\minipage{0.5\textwidth}
\centering
\includegraphics[width=\linewidth]{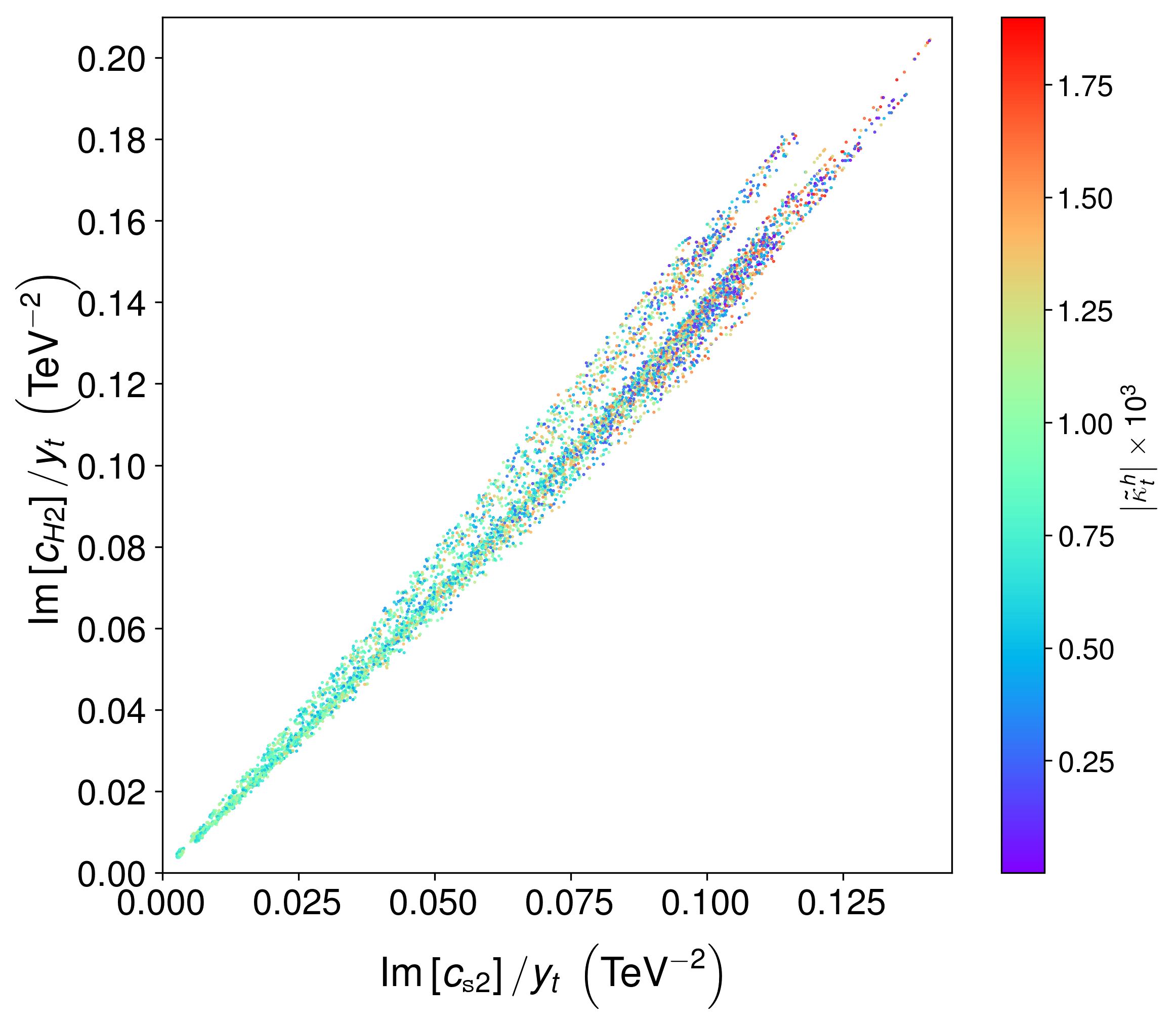}
\subcaption{}
\label{fig:ImC5s2_vs_ImC6H2_close_to_EDM_bound_with_kappa}
\endminipage\hfill
\minipage{0.5\textwidth}
\centering
\includegraphics[width=\linewidth]{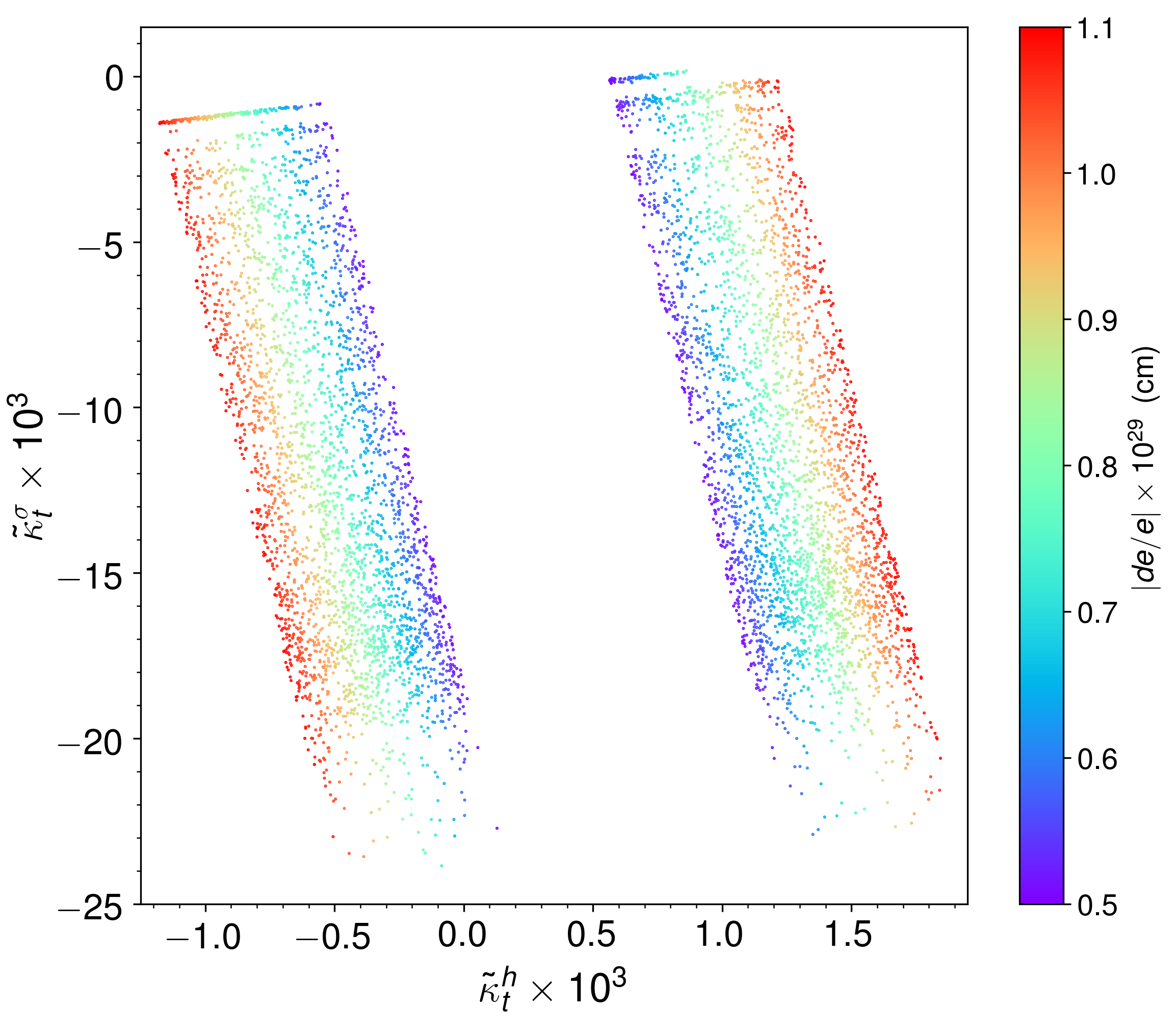}
\subcaption{}
\label{fig:kappaHiggs_vs_kappaSigma_close_to_EDM_bound}
\endminipage
\caption{{\em Values of $\Im[c^t_{\ns 1}]/y_t$, $\Im[c^t_{\ns 2}]/y_t$, $\Im[c^t_{H 2}]/y_t$, $\tilde\kappa^h_t$ and $\tilde\kappa^\sigma_t$ for the points of our \MLsM~parameter scan. The colour of the scan points markes the range of values  they give for $\tilde\kappa^h_t$ in (a), (b) and (c), and  for $\absval{d_e/e}$ in (d). The electron EDM was required to be in the range $|d_e/e|=(0.5-1.1)\times 10^{-29} \, \text{cm}$.}}
\label{fig:im_WC_close_to_EDM_bound_with_kappa}
\end{figure}

In Fig.~\ref{fig:im_WC_close_to_EDM_bound_with_kappa}, we show the results of a scan over certain regions of the parameter space of the model and impose the most relevant experimental constraints. We follow Ref.~\cite{Aguilar-Saavedra:2019ghg}, which contains a comprehensive study of the model phenomenology addressing in particular the effects of relatively light top and bottom partners.
This analysis includes the study of the modification of the $Z\bar bb$ coupling having an impact on the ratios of partial widths $R_b=\Gamma(Z\to\bar bb)/\Gamma(Z\to\text{hadrons})$, $R_c=\Gamma(Z\to\bar cc)/\Gamma(Z\to\text{hadrons})$, on the forward-backward charge asymmetry $A_{FB}^b$ and on the coupling $A^b$
from left-right forward-backward asymmetry, which have been precisely measured at LEP~\cite{ALEPH:2005ab}.  The bounds from $S$ and $T$ oblique parameters have also been considered, aside from constraints on the ML$\sigma$M scalar sector, already discussed in Sec.~\ref{SubSect.Scalar}. Reproduction of experimental values for $m_t$ and $m_b$ has been required as well. Although in our scan all those constraints are imposed at the level of real parameters, we checked that introducing the phases typically change the results by only a few percent.

The scatter plots shown in Fig.~\ref{fig:im_WC_close_to_EDM_bound_with_kappa} have been obtained for the following ranges of the parameters:
\be
\begin{gathered}
m_\sigma = 1.5 \TeV \,, \qquad\qquad \qquad
\sin^2\gamma = 0.08 \,,\\
\mo = ( 3.0 - 3.5) \TeV \,, \qquad\qquad 
\mf = ( 6.0 - 7.0 ) \TeV \,, \\ 
\mpo = ( 4.5 - 6.5 ) \TeV \,, \qquad\qquad 
\mpf = ( 7.5 - 10.0 ) \TeV \,,\\ 
\lamo = ( 5.5- 7.5 ) \TeV \, , \qquad 
\Lambda_{2,3} = ( 1.5 - 2.0 ) \TeV \, , \qquad 
\Lambda^\prime_{1,2,3} = ( 0.75 - 1.5 ) \TeV \,,\\ 
|\yt| = ( 1.00 - 1.25 )  \, , \qquad
|\ypt|=0 \, , \qquad
\alpha_{1,2} = (0 - \pi ) \, , \qquad
\alpha^\prime_{1,2} = 0 \, ,
\end{gathered}
\label{eq:scan_parameters}
\ee
Both $|\yo|$ and $|\ypo|$ where set to satify Eq.~\eqref{eq:leading_order_m_SM}, for the central experimental values for $m_t$ and $m_b$. (See some benchmark examples in Tab~\ref{tab:benchmarks}.) It has ben required that the electron EDM is in the range $|d_e/e|=(0.5-1.1)\times 10^{-29} \, \text{cm}$. The main observation is that the Wilson coefficients of different operators are strongly correlated with each other and the bounds on them are up to two orders of magnitude weaker than in the previous section. The cancellations between contributions of different operators occur naturally, as a consequence of correlations between them when they are expressed in terms of the Lagrangian parameters. No significant fine tuning of the Lagrangian parameters is needed, excluding $\alpha_2$, whose value is discussed below. In addition to the structures encoded in Eqs.~\eqref{eq:Im_WC_MLsM} the important role in the cancellations is played by the fact that in the full model the suppression scales of dim 5 and dim 6 are independent.

In Fig.~\ref{fig:kappaHiggs_vs_kappaSigma_close_to_EDM_bound} there are collected the results for the measurable parameters $\tilde\kappa_t^h$ and $\tilde\kappa_t^\sigma$. The two bands correspond to positive and negative values of $\tilde\kappa^{eff}_t$. It is interesting to observe that, contrary to the expectations based on a single operator dominance discussed in the previous section, typically $\tilde\kappa^\sigma_t\gg\tilde\kappa^h_t$, as a consequence of the large cancellations.
Interestingly, the EDM bound on the imaginary part of the $htt$ coupling in the full model is about $50\%$ weaker than in the effective approach.

\begin{figure}[h!]
\centering
\begin{subfigure}{0.45\textwidth}
\centering
\includegraphics[width=\textwidth]{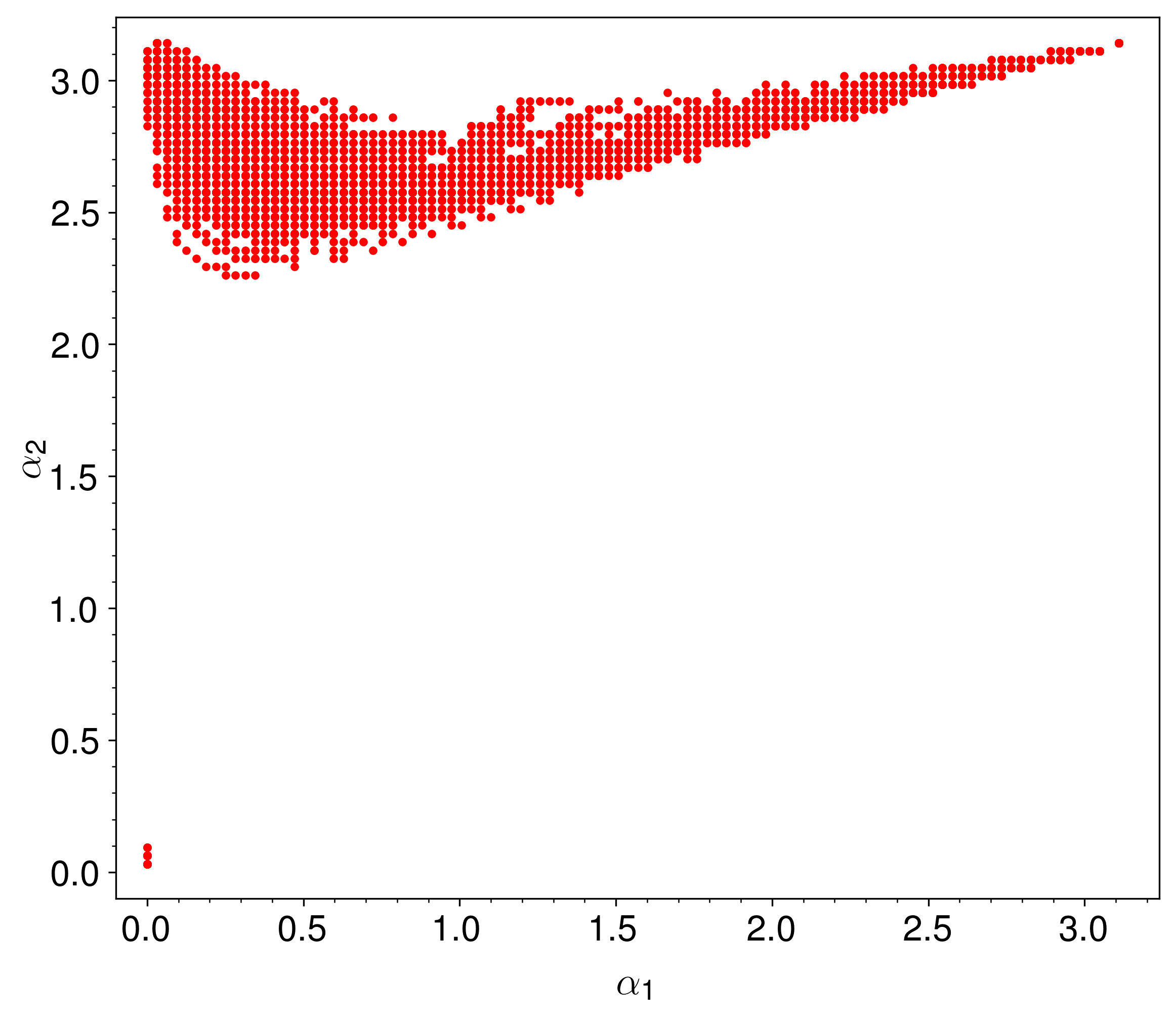}
\caption{}
\label{fig:alpha1_vs_alpha2_close_to_EDM_bound}
\end{subfigure}
\hfill
\begin{subfigure}{0.45\textwidth}
\centering
\includegraphics[width=\textwidth]{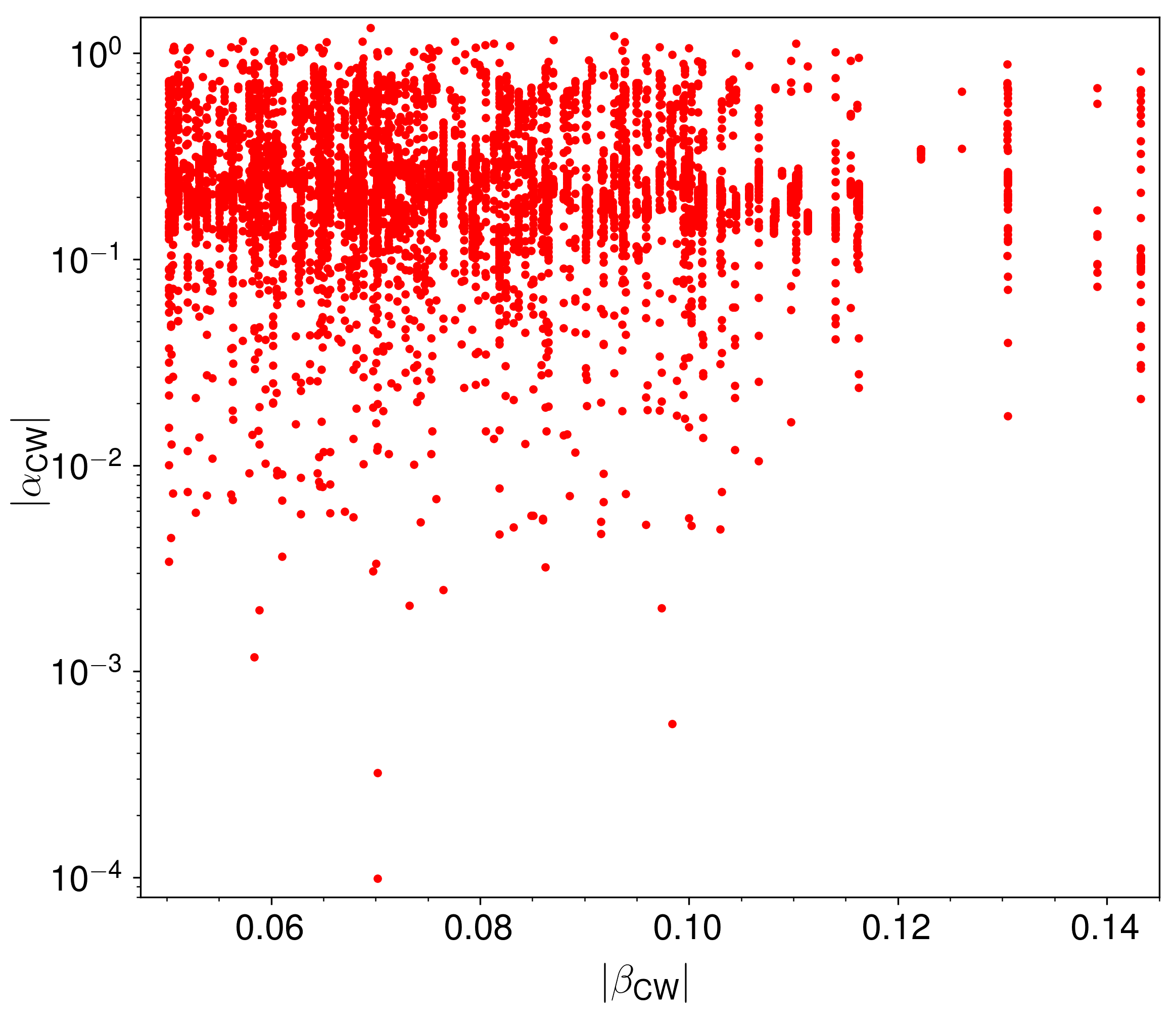}
\caption{}
\label{fig:betaCW_vs_alphaCW_close_to_EDM_bound}
\end{subfigure}
\caption{{\em Values of the CPV phases $\alpha_{1,2}$ (a) and Coleman-Weinberg parameters $\alpha_{\text{CW}}$ and $\beta_{\text{CW}}$ for the points of our \MLsM~parameter scan. The electron EDM was required to be in the range $|d_e/e|=(0.5-1.1)\times 10^{-29} \, \text{cm}$.}}
\label{fig:phases_and_CW_parameters}
\end{figure}

Fig.~\ref{fig:alpha1_vs_alpha2_close_to_EDM_bound} shows the scatter plot of the CPV Yukawa phases $\alpha_{1,2}$ compatible with $|d_e/e|=(0.5-1.1)\times 10^{-29} \, \text{cm}$, for the range of the parameters in Eq.~\eqref{eq:scan_parameters}. One can ask if such values of the CPV phases are ``natural'' or perhaps they correspond to some undesired tuning. Following traditional naturalness criterion~\cite{Barbieri:1987fn}, the amount of tuning in $\alpha_{1,2}$ for the effective coupling $\tilde\kappa^t_{h}$ can be quantified by
\begin{equation}
\Delta_{1,2}^{h} \equiv \absval{\frac{\alpha_{1,2}}{\tilde\kappa_t^{h}}\frac{\partial \tilde\kappa_t^{h}}{\partial \alpha_{1,2}}} \, .
\label{eq:tuning}
\end{equation}
Similar quantities can be defined for $\tilde\kappa_t^{\sigma}$. Tab~\ref{tab:benchmarks} collects the values of $\Delta_{1,2}^{h,\sigma}$ for some benchmarks, showing that for $\alpha_2$ a tuning of $10\%$ to $1\%$ is required in the parameter space region~\eqref{eq:scan_parameters}. This is the price for the cancellations between contributions from different effective operators.

\begin{table}[h!]
\centering 
\tabcolsep=0.1cm 
\footnotesize 
\begin{tabular}{|c|c|c|c|c|c|c|c|c|c|c|c|c|c|c|c|c|} 
\hline
& $\mo$ & $\mf$ & $\mpo$ & $\mpf$ & $\lamo$ & $\lamt$ & $\lamth$ & $\lampo$ & $\lampt$ & $\lampth$ & $|\yo|$ & $|\yt|$ & $\alpha_1$ & $\alpha_2$ & $|\ypo|$ & $|\ypt|$ \\
& $\left(\text{TeV}\right)$ & $\left(\text{TeV}\right)$ & $\left(\text{TeV}\right)$ & $\left(\text{TeV}\right)$ & $\left(\text{TeV}\right)$ & $\left(\text{TeV}\right)$ & $\left(\text{TeV}\right)$ & $\left(\text{TeV}\right)$ & $\left(\text{TeV}\right)$ & $\left(\text{TeV}\right)$ & & & & & & \\
\hline\hline
P1 & $3.4$ & $6.9$ & $5.1$ & $8.0$ & $7.1$ & $1.5$ & $2.0$ & $1.2$ & $1.1$ & $1.0$ & $2.8$ & $1.1$ & $0.03$ & $3.11$ & $1.2$ & $0$ \\
\hline
P2 & $3.1$ & $6.4$ & $6.3$ & $8.3$ & $7.1$ & $1.6$ & $1.9$ & $1.5$ & $1.3$ & $1.3$ & $2.7$ & $1.1$ & $0.22$ & $2.80$ & $1.1$ & $0$\\
\hline
P3 & $3.2$ & $6.0$ & $6.1$ & $9.1$ & $5.7$ & $1.9$ & $1.9$ & $1.1$ & $1.2$ & $1.3$ & $3.0$ & $1.0$ & $0.50$ & $2.90$ & $1.3$ & $0$\\
\hline
P4 & $3.1$ & $6.1$ & $5.1$ & $7.6$ & $6.2$ & $1.5$ & $1.7$ & $1.5$ & $1.0$ & $1.2$ & $2.9$ & $1.2$ & $0.75$ & $2.42$ & $0.8$ & $0$\\
\hline
\end{tabular}

\bigskip

\begin{tabular}{|c|c|c|c|c|c|c|c|c|} 
\hline
& $\Im[c^t_{\ns 1}]/y_t$ & $\Im[c^t_{\ns 2}]/y_t$ & $\Im[c^t_{H 2}]/y_t$ & $\tilde\kappa^h_t$ & $\tilde\kappa^\sigma_t$ & $\left( d_e/e \right)_{h}\times 10^{29}$ & $\left( d_e/e \right)_{\sigma}\times 10^{29}$ & $\left| d_e/e \right|\times 10^{29}$ \\
& $\left(\text{TeV}^{-1}\right)$ &$\left(\text{TeV}^{-2}\right)$ & $\left(\text{TeV}^{-2}\right)$ &  & & $\left(\text{cm}\right)$ & $\left(\text{cm}\right)$ & $\left(\text{cm}\right)$ \\
\hline\hline
P1 & $-0.004$ & $0.006$ & $0.008$ & $-0.0006$ & $-0.0014$ & $-0.57$ & $-0.04$ & $0.6$ \\
\hline
P2 & $-0.048$ & $0.056$ & $0.076$ & $-0.0003$ & $-0.0099$ & $-0.31$ & $-0.27$ & $0.6$ \\
\hline
P3 & $-0.061$ & $0.069$ & $0.111$ & $-0.0008$ & $-0.0118$ & $-0.70$ & $-0.32$ & $1.0$ \\
\hline
P4 & $-0.126$ & $0.140$ & $0.204$ & $0.0017$ & $-0.0226$ & $1.56$ & $-0.62$ & $0.9$ \\
\hline
\end{tabular}

\bigskip

\begin{tabular}{|c|c|c|c|c|c|c|c|c|} 
\hline
& $m_T$ & $m_B$ & $\Delta_{1}^{h}$ & $\Delta_{2}^{h}$ & $\Delta_{1}^{\sigma}$ & $\Delta_{2}^{\sigma}$ & $\abs{\alpha_{\text{CW}}}$ & $\abs{\beta_{\text{CW}}}$ \\
& $\left(\text{TeV}\right)$ & $\left(\text{TeV}\right)$ & & & & & & \\
\hline\hline
P1 & {$4.0$} & {$5.1$} & {$1.3$} & {$31.6$} & {$0.7$} & {$27.0$}  & $0.22$ & $0.05$ \\
\hline
P2 & {$3.7$} & {$6.3$} & {$12.1$} & {$149.1$} & {$0.6$} & {$1.0$}  & $0.24$ & $0.06$\\
\hline
P3 & {$3.6$} & {$6.1$} & {$4.3$} & {$121.6$} & {$0.7$} & {$2.0$}  & $0.32$ & $0.09$\\
\hline
P4 & {$3.6$} & {$5.2$} & {$7.1$} & {$85.9$} & {$0.3$} & {$4.2$}  & $0.12$ & $0.05$\\
\hline
\end{tabular}
\caption{\em In the upper panel, four different sets of input parameters are shown as examples. In all cases, $m_\sigma =1.5 \TeV $, $\sin^2\gamma=0.08$ and $\alpha^\prime_{1,2} = 0$. The corresponding imaginary parts of the Wilson coefficients, effective coefficients $\tilde\kappa^{h,\sigma}_t$ and the electron EDM are collected in the middle panel, together with the respective Higgs and singlet contributions to the later. In the lower panel are shown the resulting lightest $T$ and $B$ quark partner masses, the $\Delta_{1,2}^{h,\sigma}$ that quantify the amount of tuning in the ML$\sigma$M CPV phases and the $\abs{\alpha_{\text{CW}}}$ and $\abs{\beta_{\text{CW}}}$ parameters calculated from the Coleman-Weinberg potential. }
\label{tab:benchmarks}
\end{table}

In Tab.~\ref{tab:benchmarks} we collect the parameters and the results for a few generic points from our scan. One new important observation is that in the full model the  mass scale of new fermions compatible with the electron EDM bound is around 3 TeV, for the ${\cal O}(1)$ phases of the Yukawa phases in the lagrangian. This is the effect of the already discussed cancellations.

Finally, the second question we discuss in this section is this: can the scalar potential parameter range shown in Fig.~\ref{fig:sg2_VS_ms} be naturally obtained from radiative corrections induced by the soft $SO(5)$ breaking terms in the ML${\sigma}$M Lagrangian Eq.~\eqref{ScalarPotential} (respecting the naturalness cut-off $\Lambda_{nat}\approx {\cal O}(10)f$ for the heaviest exotic fermion mass). 

The $\alpha_{\text{CW}}$ and $\beta_{\text{CW}}$ parameters arising form the Coleman-Weinberg (CW) one-loop potential are given by

\begin{equation}
\begin{split}
& \left|\alpha_{\text{CW}}\right| \equiv \left|\frac{1}{64 \, \pi^2} \frac{d_2}{f^3}\right| \, , \quad 
\left|\beta_{\text{CW}}\right| \equiv \left|\frac{1}{64 \, \pi^2} \frac{d_3}{f^2}\right| \, ,
\end{split}
\end{equation}
where $d_{2,3}$ have been calculated in Ref.~\cite{Feruglio:2016zvt} (see Ref.~\cite{Merlo:2017sun} for a different treatment)
\begin{align}
d_2  &= 4 \, \Lambda_2 \Lambda_3 \left(M_1 \left|\yo\right| \cos\alpha_1 + M_5 \left|\yt\right|  \cos\alpha_2 \right) +
4 \, \Lambda^\prime_2 \Lambda^\prime_3 \left(M^\prime_1 \left|\ypo\right| \cos\alpha^\prime_1 + M^\prime_5 \left|\ypt\right| \cos\alpha^\prime_2  \right)
\, ,\nn \\
d_3  &= 2 \, \left|\yo\right|^2 \Lambda_2^2 - \left|\yt\right|^2 \Lambda_1^2 + 2 \, \left|\ypo\right|^2 \Lambda_2^{\prime 2} - \left|\ypt\right|^2 \Lambda_1^{\prime 2} \, .
\end{align}
Recall that $\ypo$ and $\ypt$ were set to be real since the bottom role in the EDM is subleading.

The results of a scan over the parameters of the Lagrangian are shown in Fig.~\ref{fig:betaCW_vs_alphaCW_close_to_EDM_bound}. There can be a good agreement with the values of the parameters indicated in Eq.~\ref{eq:scalar_param_values}.

The largeness of the $\yo$ parameter at the average scale of the decoupling of the heavy fermions may raise the question about the appearance of Landau poles. However, that scale is also similar to  $\Lambda_{nat}$ that is the natural cut-off for the \MLsM at which  a (still perturbative) completion of the theory is expected.  The running of the couplings above $\Lambda_{nat}$ would depend on that completion so the question about the Landau pole is beyond the \MLsM itself. However, just for controlling the effects of the possible short running in the \MLsM (given the uncertainty of the scale at which the value of $\yo$ should be assigned to), we have checked that
with the 1-loop beta-function for $\yo$ calculated for our spectrum, the corresponding Landau pole would arise at $\sim13\TeV$ with  $\yo(M_5=6 \TeV)=3$.

\boldmath
\section{Electroweak Phase Transition}
\label{Sect.Phase}
\unboldmath

The \MLsM is an example of extension of the SM with an additional scalar. Various extensions of the SM with additional scalars have been widely studied to make the EW phase transition first order and strong~\cite{Espinosa:2007qk,Profumo:2007wc,Ashoorioon:2009nf,Espinosa:2011ax,Espinosa:2011eu,Chung:2012vg,Cline:2013gha,Alanne:2014bra,Profumo:2014opa,Curtin:2014jma,Chala:2016ykx,Bruggisser:2018mus,Bruggisser:2018mrt}. This is one of the necessary conditions for a successful EW baryogenesis~\cite{Sakharov:1967dj,Kuzmin:1985mm,Riotto:1999yt}. A natural question is if the \MLsM can manifest such strong phase transition.

To analyze the phase transitions a model exhibits, the effective potential of the scalar sector at finite temperature needs to be studied. This is calculated using the methods presented in the Appendix~\ref{App:FiniteTemp}. 
At finite temperature, the quadratic terms of the scalar potential receive thermal contributions, so above some critical temperature $T_c$ the EW symmetry is restored.

The transition between the two phases is first order if at $T_c$, there is a barrier in the effective potential between the two degenerate minima: the EW symmetric minimum at $\nh=0$, and the EW breaking minimum at $\nh=v_{\nh,c}$. Then, the phase transition occurs through bubble nucleation: at $T>T_c$ the plasma is in the EW symmetric phase, and at $T<T_c$, bubbles in the EW breaking phase start to grow and percolate. This is a highly out-of-equilibrium process, and an excess of baryon number can be generated in the bubble walls~\cite{Cohen:1993nk, Morrissey:2012db}.
To explain the observed baryon number excess, the phase transition must also be strong enough: $v_{\nh,c}/T_c>0.6-1.6$ (see Refs.~\cite{Patel:2011th,Curtin:2014jma}).

In the SM the phase transition has a cross over character (the Higgs gets a VEV smoothly without nucleation), but extensions with scalar fields can change this in a number of ways~\cite{Chung:2012vg}. 
If the extra-scalar VEVs are invariant during the EW phase transition and only renormalizable interactions are allowed, the tree level potential cannot have a barrier~\cite{Espinosa:2011ax}, but loop and thermal contributions of the extra scalars can generate it~\cite{Espinosa:2007qk}.
Those models typically require sizeable couplings to make the phase transition strong.
A second class of models are those ones where the extra-scalar VEVs change during the EW phase transition. 
In this case, the tree level potential may have a barrier assuming only renormalisable interactions~\cite{Espinosa:2011ax}.

Models with an extra scalar with a $Z_2$ symmetry, and a vanishing VEV at $0$ temperature are considered in~\cite{Cline:2013gha,Alanne:2014bra,Curtin:2014jma}. If the VEV of the extra scalar also vanishes in the EW symmetric phase, it corresponds to the first case explained above. The second case can also be realized with a two-step phase transition if the extra scalar gets a VEV before the EW symmetry is broken.
In CH models, the second case can be implemented with help of extra pseudo-Goldstone bosons in non-minimal models~\cite{Espinosa:2011eu,Chala:2016ykx} or the dilaton~\cite{Bruggisser:2018mus,Bruggisser:2018mrt}.

In the \MLsM there is an extra scalar $\sigma$ which arises mainly from the radial mode of a scalar 5-plet, whose VEV $f$ breaks spontaneously the global $SO(5)$ symmetry. 
As explained in detail below, although the VEV of $\ns$ does change during the EW phase transition, the underlying $SO(5)$ global symmetry of the potential does not allow for a tree-level potential with a barrier.
Several effects also suppress the quantum and thermal contributions of $\sigma$ to create a barrier: very strong experimental upper bound on the $\nh-\ns$ mixing and the hierarchy $f/v_{\nh}\gg1$ at zero temperature. The phase transition is by far too weak to be consistent with electroweak baryogenesis.

\subsection{Leading Temperature Corrections}

The first step is to analyse the effective potential including only leading high temperature corrections (see Appendix~\ref{App:HighTemp}). This approximation can neglect important contributions from higher orders, but it gives a first analytic approach to the problem. In the next subsection the full numerical analysis will be performed.
The effective potential in this approximation~\eqref{HTEffpot} has the same functional form as the tree level potential~\eqref{ScalarPotential} with some temperature-dependent coefficients (up to constant terms):
\begin{equation}
V_{\mathrm{eff},0}(\nh,\ns)=\lambda\left(\nh ^2+\ns^2-f(T)^2\right)^2+\alpha(T) f(T)^3 \ns-\beta(T) f(T)^2 \nh^2, \label{VTHTL}
\end{equation}
where
\begin{align}
f(T)^2&=f^2-\frac{7}{12}T^2,\\
\alpha(T)&=\alpha \frac{f^3}{f(T)^3},\\
\beta(T)&=\frac{1}{f(T)^2}\left(f^2\beta -\frac{T^2}{8v_{\nh}^2}(m_Z^2+2m_W^2+2m_t^2)  \right).
\end{align}
The possible first order phase transitions can be studied looking for the parameters that give a potential~\eqref{VTHTL} with two degenerate minima. Then, evolving the parameters from $T=T_c$ to $T=0$, the zero temperature tree level parameters can be found~\cite{Espinosa:2011ax}.
However, if $\alpha \neq 0$, the potential~\eqref{VTHTL} cannot even have two local minima if one of them breaks the EW symmetry.
This potential has a stationary point with $v_{\nh}\neq 0$ if and only if the condition
\begin{equation}
f(T)^2\,\alpha(T)^2 < \, 4 \,f(T)^2\,\beta(T)^2\left(1+\frac{\beta(T)}{2\lambda} \right) \label{CondEWSt}
\end{equation}
is satisfied. Notice that, assuming $f^2>0$, this condition is the right inequality~\eqref{TheoContraintsab} . The stationary point is unique (up to $SO(4)$ transformations) and it is a minimum if and only if
\begin{equation}
f(T)^2 \beta(T) >0.\label{CondEWMin}
\end{equation}
If there is a second minimum, it has to be located along $\nh=0$. However, once the inequalities~\eqref{CondEWSt} and~\eqref{CondEWMin} are satisfied, all roots of $\partial V_{\mathrm{eff}}/\partial \ns$ along $\nh=0$ are contained in the interval where $\partial^2 V_{\mathrm{eff}}/\partial \nh ^2$ is negative. All stationary points along $\nh=0$ are then either maxima or saddle points.
Therefore, in this approximation all phase transitions are second order (VEVs evolve continuously with the temperature).
The absence of a tree level barrier is related to the flat direction of the $SO(5)$ potential, which is too weakly perturbed.\footnote{It remains to be seen if adding more $SO(5)$-breaking terms can give us a strong first order phase transition suitable for baryogenesis.}

The symmetry-breaking pattern as the temperature decreases is as follows. At very high temperatures, both $f(T)^2$ and $f(T)^2\beta(T)$ are negative, so the quadratic terms for $\nh$ and $\ns$ at the origin are positive, and the $SO(5)$ symmetry is not spontaneously broken.
The $f(T)^2$ is negative for temperatures higher than
\begin{equation}
T_{SO(5)}=\sqrt{\frac{12}{7}}f.
\end{equation}
When the conditions~\eqref{CondEWSt} and~\eqref{CondEWMin} are satisfied, the EW symmetry is broken. The saturating point of~\eqref{CondEWSt} once~$f^2(T)\beta(T)\geq 0$ gives the EW critical temperature $T_{EW}$.
If the zero-temperature parameters satisfy
\begin{align}
\beta&<\frac{\alpha^{2/3}\lambda^{1/3}}{2^{1/3}}+\frac{3}{14 v_{\nh}^2}(m_Z^2+2 m_W^2+2 m_t^2)\nonumber \\
&= \frac{\alpha^{2/3}\lambda^{1/3}}{2^{1/3}}+0.287,
\label{CondHTEW}
\end{align}
then $T_{EW}<T_{SO(5)}$. At $T_{SO(5)}$, the $SO(5)$ symmetry is broken spontaneously but the EW symmetry is preserved because of the explicit $SO(5)$ breaking. The VEV $v_{\ns}(T)$ starts to receive increasing contributions from this breaking while $v_{\nh}(T)$ keeps vanishing. The EW symmetry is later broken at a lower temperature.
If the condition~\eqref{CondHTEW} is not satisfied, then $T_{EW}\geq T_{SO(5)}$ and the EW symmetry is broken at the same time as the $SO(5)$ symmetry, before $f(T)^2$ becomes positive. However, in that case, $f<v_{\nh}$ ($T_{EW}$ is still at the EW scale), and it corresponds to a very tiny region in the allowed parameter space very close to the $f^2=0$ limit.

\subsection{Full Analysis}

Although leading temperature contributions do not trigger a first order phase transition, it is well known that higher orders can generate a barrier from the bosonic contribution in the Higgs potential if the coupling between the Higgs and the scalar is large enough. 
A natural question is whether the loop-contribution of $\ns$ in the potential is enough to generate such a barrier in the EW phase transition.

\begin{figure}
\centering
\includegraphics[width=0.7\textwidth]{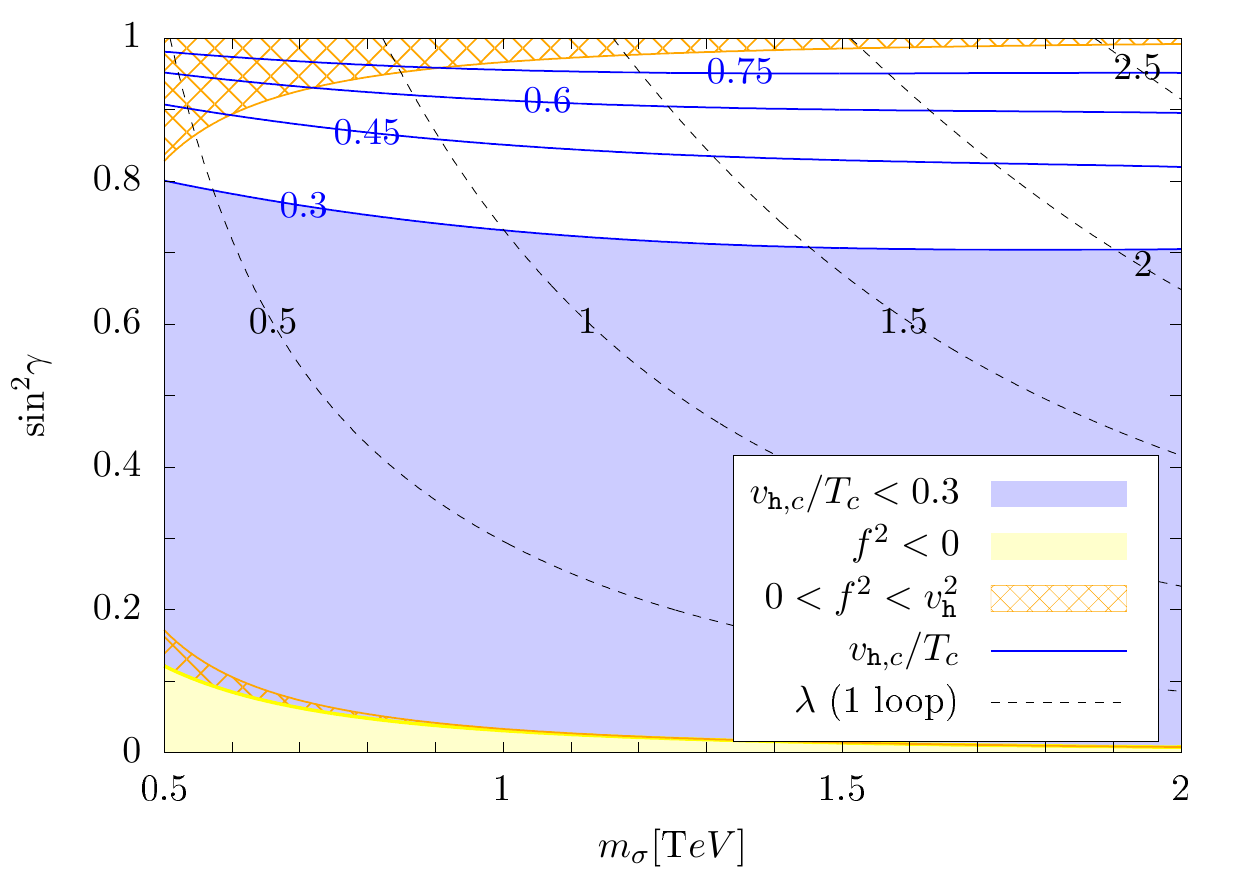} 
\caption{\em Strength of the EW phase transition ($v_{\nh,c}/T_c$) as function of the singlet mass ($m_{\sigma}$) and the mixing angle ($\sin^2 \gamma$). The blue solid lines correspond to $v_{\nh,c}/T_c=0.3,\,0.45,\,0.6,\,0.75$. In addition, the black dashed lines depict $\lambda=0.5,\,1,\,1.5,\,2,\,2.5$, calculated with the one-loop effective potential. According to the criterion of Ref.~\cite{Curtin:2014jma}, in the region where $\lambda\gtrsim 1.5$, the one-loop analysis of the phase transition breaks down due to higher loop corrections.}
\label{fig:PTStrength}
\end{figure}

If we assume $v_{\nh}\ll f$ and $T_{EW}\ll T_{SO(5)}$, the VEV of $\ns$ will be approximately constant in the EW phase transition. If we perform a shift on $\ns$ to make $v_{\ns}=0$, the situation becomes similar to the one-step phase transition discussed in Ref.~\cite{Curtin:2014jma}. There, the EW phase transition for the SM plus a scalar that does not develop a VEV (before and after the phase transition) was studied.
By analogy, in our case, the operator responsible for the barrier in the potential $V(\nh,v_{\ns})$ is $\nh^2 (\ns-v_{\ns})^2$. In the \MLsM, the coupling of this operator is $2\lambda$.
Translating directly the results of Ref.~\cite{Curtin:2014jma} to the \MLsM case one obtains that, if the mass of the singlet is in the range $0.5 - 1\,$TeV, the coupling must be $\lambda \gtrsim 0.75-1.5$ for the phase transition to be relatively strong $v_{\nh,c}/T_c \gtrsim 0.6$. However, these values require a Higgs-singlet mixing $\sin^2 \gamma > 0.7$ which is completely excluded (see Eq.~\eqref{sin2gammaBound}). This suggests that a strong first order phase transition is unlikely in the ML$\sigma$M.

To confirm the previous argument, the full numerical calculation has been performed, including all finite temperature corrections up to one-loop and ring corrections according to Appendix~\ref{App:FiniteTemp}. The results are presented in Fig.~\ref{fig:PTStrength}. For the non-excluded low mixing angles ($\sin^2 \gamma \lesssim 0.09$), the contribution of the singlet to the barrier is absolutely negligible. Because of this, the full range of the mixing angle is depicted in order to access to the range of parameters where the phase transition becomes strong.
Indeed, only for completely excluded mixing angles $\sin^2 \gamma \gtrsim 0.9$, the system manifests a strong phase transition ($v_{\nh,c}/T_c \gtrsim 0.6$).

\boldmath
\section{Conclusions}
\label{Sect.Conclusions}
\unboldmath
The SMEFT Lagrangian extended by operators built in terms of an additional electroweak scalar field $\ns$ which has a non-vanishing vacuum expectation value is an EFT approximation to a class of perturbative theories with the Higgs boson appearing as a Goldstone boson of a spontaneously broken global symmetry. In particular it contains dim 5 and  dim 6 operators.

We have analysed the electron EDM bounds on the pseudoscalar Yukawa couplings of the top quark in such an EFT. The contributions from the new operators, if uncorrelated, leads to much stronger bounds on the scale of new physics than in SMEFT. This is because they depend on the large VEV of the field $\ns$.

The ML$\sigma$M is an example of a model with the radial mode of a scalar $\ns$ breaking a global symmetry in the spectrum. The Wilson coefficients of the EFT operators are calculable in terms of the original parameters of the Lagrangian and strongly correlated. Those correlations weaken the bounds on the scale of new states in the model by more than two orders of magnitude compared to the ``uncorrelated" EFT approach. The price is some fine-tuning
in the phases of the Yukawa couplings in the Lagrangian, of the order of $10\%$ to $1\%$.

Thus, the model provides an explicit example that in UV complete models the lightest NP mass eigenstates can be in the range  of $\Lambda={\cal O}(1 \TeV)$ even for {\cal O}(1) phases of the Lagrangian parameters, and remaining consistent with the EDM bound on the complex top Yukawa coupling. This is in contrast  to the  EFT approach if the correlations between Wilson operators are neglected.

The UV complete model allows for an unambiguous calculation of the electroweak phase transition. In the allowed
parameter range (Higgs and scalar singlet mixing), it turns out to be by far too weak for a successful baryogenesis.

\section*{Acknowledgements}
We thank J.~R. Espinosa, M. Olechowski, K. Sakurai, J. van de Vis and G.~A. White for fruitful discussions. 
S.P. is grateful to Anna Lipniacka (Bergen University) for discussions on the prospects of measuring CPV in the tau Yukawa coupling.

 J.A.G. and L.M. acknowledge partial financial support by the Spanish MINECO through the Centro de excelencia Severo Ochoa Program under grant SEV-2016-0597, by the Spanish ``Agencia Estatal de Investigac\'ion''(AEI) and the EU ``Fondo Europeo de Desarrollo Regional'' (FEDER) through the projects FPA2016-78645-P and PID2019-108892RB-I00/AEI/10.13039/501100011033.
 J.A.G. and L.M. also acknowledge support from the European Union's Horizon 2020 research and innovation programme under the Marie Sk\l odowska-Curie grant agreement No 860881-HIDDeN.
J.M.L. acknowledges financial support from the Polish National Science Center under the Beethoven series grant number DEC-2016/23/G/ST2/04301, and from the European Research Council (ERC) under the European Union's Horizon 2020 research and innovation program under grant agreement 833280 (FLAY).
L.M. acknowledges partial financial support by the Spanish MINECO through the ``Ram\'on y Cajal'' programme (RYC-2015-17173).
The research of S.P. has received funding from the Norwegian Financial Mechanism for years 2014-2021, grant nr 2019/34/H/ST2/00707. 

\appendix

\section{Finite Temperature Effective Potential}
\label{App:FiniteTemp}

In order to study possible phase transitions, we need to calculate the finite-temperature effective potential for the scalar sector. We will work at one loop including ring corrections:
\begin{equation}
V_{\mathrm{eff}}(T;\phi)=V(\phi)+V_{\mathrm{CW}}(\phi)+V_{\mathrm{T}}(T;\phi)+V_{\mathrm{r}}(T;\phi).
\end{equation}
Here, $V$ is the tree level scalar potential~\eqref{ScalarPotential}, $V_{\mathrm{CW}}$ is the Coleman-Weinberg correction (one-loop zero temperature correction), $V_{\mathrm{T}}$ is the one-loop finite temperature correction, and $V_{\mathrm{r}}$, the ring correction. It is convenient to keep all the components of the scalar $\phi=(\pi_1,\pi_2,\pi_3,\nh,\ns)$ because in the calculation of every piece of the effective potential we have to take into account the Goldstone bosons $\vec \pi$. Once computed, we can always take the unitary gauge $\vec\pi=0$.

The Coleman-Weinberg correction~\cite{Coleman:1973jx} in $\overline{\textrm{MS}}$ is given by
\begin{equation}
V_{\mathrm{CW}}(\phi)=\sum_{i=(\phi,f,V)} (-1)^{F_i}\frac{n_i}{64 \pi^2}\Tr\left[\mathrm{M}_i^4\left(\log \frac{\mathrm{M}_i^2}{\mu^2_{\overline{\textrm{MS}}}}-C_i \right) \right].
\end{equation}
Here, $i$ runs over the different spin sectors $i=(\phi, f,V)$ (scalars, fermions, and gauge bosons), $F_i$ is the fermion number ($1$ for fermions, $0$ for bosons), and $\mathrm{M}_i$ is the mass matrix for the sector $i$ for a given VEV for $\phi=(\pi_1,\pi_2,\pi_3,\nh,\ns$). For the scalar and vector sectors (in gauge basis):
\begin{align}
\left(\mathrm{M}^2_{\phi}\right)_{ab}=&\frac{\partial^2 V}{\partial \phi^{a}\partial \phi^b},\\
\mathrm{M}^2_{V}=&\frac{(\nh^2+\vec\pi^2)}{v_{\nh}^2}
\begin{pmatrix}
m_W^2 &0 & 0 & 0 \\
0 & m_W^2 & 0 & 0 \\
0 & 0 & m_W^2 & -m_W\sqrt{m_Z^2-m_W^2}\\
0 & 0 & -m_W\sqrt{m_Z^2-m_W^2} & m_Z^2-m_W^2
\end{pmatrix}.
\end{align}
In the fermion sector, we neglect the contributions from the effective operators of dimension 5 or higher of Tab.~\ref{tabops}, and the contribution from all SM fermions except the top. The fermion sector reduces then to 
\begin{equation}
\mathrm{M}^2_f=\frac{m^2_t}{v_{\nh}^2}(\nh^2+\vec\pi^2).
\end{equation}
Lastly,
\begin{align}
n_i=&(1,12,3),\\
C_i=&\left(\frac{3}{2},\frac{3}{2},\frac{5}{2}\right),
\end{align}
where we have included the color factor in $n_f$. For the calculations in this work we set the renormalization scale to be $\mu_{\overline{\textrm{MS}}}=v_{\nh}$.

The one-loop finite temperature correction is~\cite{Dolan:1973qd,Weinberg:1974hy}
\begin{equation}
V_{\mathrm{T}}(T;\phi)= \sum_{i=(\phi,f,V)} (-1)^{F_i}\frac{n_i T^4}{2\pi^2}\Tr J_{F_i}(\mathrm{M}^2_i/T^2),
\end{equation}
where $J_0$ and $J_1$ are the thermal bosonic and fermionic functions:
\begin{align}
J_{F_i}(y^2)= \int_0^{\infty} \mathrm{d} x\,x^2\log\left[1-(-1)^{F_i} e^{-\sqrt{x^2+y^2}}  \right].\label{ThermalFunct}
\end{align}

Finally, if $T\gg M_i$, the so-called ring diagrams, which are bosonic multi-loop diagrams, can give important contributions due to large $T/\mu_{\overline{\textrm{MS}}}$ ratios, so they have to be resummed~\cite{Delaunay:2007wb}. For this we use the truncated full dressing method~\cite{Curtin:2016urg}: 
\begin{equation}
V_{r}(T;\phi)=\sum_{i=(\phi, V )} \frac{T}{12\pi} \Tr \left[\mathrm{M}_i^3-(\mathrm{M}_i^2+\Pi_i(T))^{3/2}   \right],
\end{equation}
where $\Pi_i$ is zero momentum thermal self-energies, that, in the high temperature limit are
\begin{align}
(\Pi_{\phi}(T))_{ab}=&\sum_{i=(\phi,f,V)} T^2\frac{n_i^{\prime} }{24}  \frac{\partial^2\, \Tr\mathrm{M}^2_i}{\partial \phi^a \partial\phi^b},\\
\Pi_{V}(T)=&\frac{22}{3}\frac{T^2}{v_{\nh}^2}\mathrm{diag}(m_W^2,m_W^2,m_W^2,m_Z^2-m_W^2),
\end{align}
where
\begin{equation}
n_i^{\prime}=(1,6,3).
\end{equation}
Notice that both self-energies are $\phi$-independent: for the scalar self-energy, the squared mass matrix depends at most quadratically on the fields $\phi^a$.

\subsection{High Temperature Approximation}
\label{App:HighTemp}

Expanding in the high temperature limit, the leading finite-temperature correction comes at order $O(T^2)$ from the Taylor expansion of the thermal functions~\eqref{ThermalFunct} around $y^2=0$. Including only these terms, the finite-temperature effective potential is
\begin{equation}
V_{\mathrm{eff},0}(T;\phi)=V(\phi)+\sum_{i=(\phi,f,V)}T^2\frac{n_i^{\prime}}{24} \Tr \mathrm{M}^2_i.\label{HTEffpot}
\end{equation}
In this approximation, only the quadratic and linear terms of the potential receive finite temperature contributions.


\footnotesize

\bibliography{biblio}{}
\bibliographystyle{BiblioStyle}

\end{document}